\documentclass[10pt]{iopart}
\usepackage{tikz}
\usetikzlibrary{arrows}
\usetikzlibrary{positioning,backgrounds}
\usetikzlibrary{calc,shadings}
\usetikzlibrary{positioning}
\usetikzlibrary{fit}
\usetikzlibrary{arrows.meta}
\usepackage{iopams}  
\usepackage{graphicx}
\usepackage{array}

\usepackage{tikz}
\usetikzlibrary{shapes.geometric, arrows, positioning}

\tikzstyle{startstop} = [rectangle, rounded corners, minimum width=3cm, minimum height=1cm,text centered, draw=black, fill=red!30]
\tikzstyle{io} = [trapezium, trapezium left angle=70, trapezium right angle=110, minimum width=3cm, minimum height=1cm, text centered, draw=black, fill=blue!30]
\tikzstyle{process} = [rectangle, minimum width=3cm, minimum height=1cm, text centered, text width=3cm, draw=black, fill=orange!30]
\tikzstyle{decision} = [diamond, minimum width=3cm, minimum height=1cm, text centered, draw=black, fill=green!30]
\tikzstyle{arrow} = [thick,->,>=stealth]
\tikzstyle{line} = [draw, thick, -latex']

\tikzstyle{block01} = [rectangle, rounded corners, minimum width=10cm, minimum height=2.2cm,text centered, node distance=1cm, draw=black, fill=red!30]
\tikzstyle{block02} = [rectangle, rounded corners, minimum width=6cm, minimum height=2cm, node distance=4cm, text centered, draw=black, fill=blue!30]
\tikzstyle{block03} = [rectangle, rounded corners, minimum width=5cm, minimum height=2cm, node distance=4cm, text centered, draw=black, fill=blue!30]
\tikzstyle{block04} = [rectangle, rounded corners, minimum width=10cm, minimum height=3cm,node distance=5cm, text centered, draw=black, fill=green!30]
\tikzstyle{block05} = [rectangle, rounded corners, minimum width=10cm, minimum height=3cm,node distance=5cm, text centered, draw=black, fill=green!30]

\tikzstyle{block07} = [rectangle, rounded corners, minimum width=10cm, minimum height=3cm,node distance=5cm, text centered, draw=black, fill=blue!10]
\tikzstyle{block08} = [rectangle, rounded corners, minimum width=10cm, minimum height=3cm,node distance=5cm, text centered, draw=black, fill=green!20]

\tikzstyle{startstop2} = [rectangle, rounded corners, minimum width=4cm, minimum height=1cm,text centered, draw=black, fill=red!30]
\tikzstyle{io2} = [trapezium, trapezium left angle=70, trapezium right angle=110, minimum width=3cm, minimum height=1cm, text centered, draw=black, fill=blue!30]
\tikzstyle{process2} = [rectangle, minimum width=5cm, minimum height=1cm, text centered, text width=5cm, draw=black, fill=orange!30]
\tikzstyle{process3} = [rectangle, minimum width=3cm, minimum height=1cm, text centered, text width=3cm, draw=black, fill=orange!30]
\tikzstyle{decision2} = [diamond, minimum width=3cm, minimum height=1cm, text centered, draw=black, fill=green!30]
\tikzstyle{arrow2} = [thick,->,>=stealth]
\tikzstyle{line2} = [draw, thick, -latex']

\usepackage{enumitem}
\usepackage{hyperref}
\usepackage{color,soul}
\hypersetup{
    colorlinks=true,
    linkcolor=blue,
    filecolor=blue,      
    urlcolor=blue,
    citecolor=blue,
    pdftitle={Overleaf Example},
    pdfpagemode=FullScreen,
    }
\usepackage{xcolor}
\usepackage{tikz}
\usepackage{comment}
\usepackage{color}
\definecolor{blue}{rgb}{0.1,0.1,0.8}
\definecolor{magenta}{rgb}{1,0,1}
\definecolor{red}{rgb}{1,0,0}

\newcommand{\x}{{\rm x}}

\newcommand{\K}{\mathcal{K}}
\newcommand{\R}{\mathbb{R}}
\newcommand{\N}{\mathbb{N}}
\newcommand{\Z}{\mathbb{Z}}

\newcommand{\eq}{\begin{equation}}
\newcommand{\eqend}{\end{equation}}
\newcommand{\eqn}[1]{(\ref{#1})}

\newcommand{\PD}[2]{\frac{\partial#1}{\partial#2}}

\newcommand{\iv}{{\bf i}}
\newcommand{\jv}{{\bf j}}
\newcommand{\pv}{{\bf p}}
\newcommand{\rv}{{\bf r}}

\newcommand{\Pv}{{\bf P}}
\newcommand{\mv}{{\bf m}}

\newcommand{\Fv}{{\bf F}}

\newcommand{\Ev}{{\bf E}}

\newcommand{\sv}{{\bf s}}

\newcommand{\Lv}{{\bf L}}
\newcommand{\Av}{{\bf A}}
\newcommand{\xv}{{\bf x}}
\newcommand{\yv}{{\bf y}}

\newcommand{\Bv}{{\bf B}}

\newenvironment{proof}{\paragraph{Proof:}}{\hfill$\square$}

\newcounter{algo}[section] 
\makeatletter
\renewcommand{\p@algo}{\thesection.}
\makeatother
\newenvironment{algorithm}[1]{\vspace{1mm}\raggedright\refstepcounter{algo}\hrule\hrule\vspace{0.5mm}\noindent\textbf{Algorithm~\thesection.\thealgo} \ttfamily#1\vspace{0.5mm}\hrule}{\raggedright\hrule\hrule\vspace{1mm}\par}
\usepackage{array}
\makeatletter
\def\hlinewd#1{%
\noalign{\ifnum0=`}\fi\hrule \@height #1 %
\futurelet\reserved@a\@xhline}
\newcolumntype{"}{@{\hskip\tabcolsep\vrule width 1pt\hskip\tabcolsep}}
\makeatother

\definecolor{lime}{HTML}{A6CE39}
\DeclareRobustCommand{\orcidicon}{%
	\begin{tikzpicture}
	\draw[lime, fill=lime] (0,0) 
	circle [radius=0.16] 
	node[white] {{\fontfamily{qag}\selectfont \tiny ID}};
	\draw[white, fill=white] (-0.0625,0.095) 
	circle [radius=0.007];
	\end{tikzpicture}
	\hspace{-2mm}
}

\foreach \x in {A, ..., Z}{%
	\expandafter\xdef\csname orcid\x\endcsname{\noexpand\href{https://orcid.org/\csname orcidauthor\x\endcsname}{\noexpand\orcidicon}}
}

\begin{document}

\title[Wigner transport in linear electromagnetic fields]{Wigner transport in linear electromagnetic fields}

\author{C Etl\orcidA{}, M Ballicchia\orcidB{}, M Nedjalkov\orcidC{} and \\J Weinbub\orcidD{}}

\address{Institute for Microelectronics, TU Wien, Gu{\ss}hausstra{\ss}e 27-29/E360, 1040 Vienna, Austria}
\ead{etl@iue.tuwien.ac.at}
\vspace{10pt}
\begin{indented}
\item[]July 2023
\end{indented}

\begin{abstract}
Applying a Weyl-Stratonovich transform to the evolution equation of the Wigner function in an electromagnetic field yields a multidimensional gauge-invariant equation which is numerically very challenging to solve. In this work, we apply simplifying assumptions for linear electromagnetic fields and the evolution of an electron in a plane (two-dimensional transport), which reduces the complexity and enables to gain first experiences with a gauge-invariant Wigner equation. We present an equation analysis and show that a finite difference approach for solving the high-order derivatives allows for reformulation into a Fredholm integral equation. The resolvent expansion of the latter contains consecutive integrals, which is favorable for Monte Carlo solution approaches.  To that end, we present two stochastic (Monte Carlo) algorithms that evaluate averages of generic physical quantities or directly the Wigner function. The algorithms give rise to a quantum particle model, which interprets quantum transport in heuristic terms.
\end{abstract}


\maketitle

\section{Introduction}

The analysis of charged quantum particles in electromagnetic fields is, among others, particularly important to nanoelectronics~\cite{kluksdahl1988quantum, arnold1989electromagnetic, chang2008berry, duque2012response,  iafrate, Cepellotti2021, ibarra2022dirac, Iafrate2022}. The established  Wigner formulation of quantum mechanics~\cite{wigner1932} (see recent reviews~\cite{Weinbub2018,Weinbub2022} and book~\cite{Ferry2018}) defines the Wigner function by applying the Weyl transform to the density matrix~\cite{tatarskiui1983wigner}:
\begin{equation}
    f_{\rm w}(\pv,\xv)=\int \frac{\rmd\sv}{(2\pi\hbar)^3}\e^{-\frac{\rmi}{\hbar}\sv\cdot\pv}\rho(\xv+\frac{\sv}{2},\xv-\frac{\sv}{2})
\label{wf}
\end{equation}
The density matrix $\rho$ of a pure state is defined from the solution $\psi$ of the Schrödinger equation   as $\rho(\xv,\yv)=\psi(\xv)\psi^*(\yv)$ and depends on two position variables. (\ref{wf}) is a transformation from the position space to the phase space, i.e., $f_{\rm w}$ is a function of the momentum $\pv$ and the position $\xv$. 
%The Wigner function is a quasi-probability density, i.e.,it is normed, but not non-negative and can be used to evaluate the expecation value of a physical quantity of a quantum mechanical system.
The evolution equation for the Wigner function is obtained by applying the Weyl transform to 
the von Neumann equation  $\rmi\hbar\PD{}{t}\hat\rho=[\hat H,\hat\rho]_-:=\hat H\hat\rho-\hat\rho \hat H,$
%(itself obtained from  the Schrödinger equation), which in operator form is $ \rmi\hbar\PD{}{t}\hat\rho=[\hat H,\hat\rho]_-:=\hat H\hat \hat\rho-\hat\rho \hat H,$
%for the evolution of $\rho$, determined by the commutator 
with the Hamiltonian $\hat H=\frac{1}{2m}\left(-\rmi\hbar\nabla\right)^2+V(\rv)$~\cite{Hall2013}. The potential energy $V$ defines a central quantity of the standard theory, namely the Wigner potential:
\begin{equation}
V_{\rm w}(\pv,\xv)=\frac{1}{(2\pi\hbar)^3}\int \frac{\rmd\sv}{\rmi\hbar}\e^{-\frac{\rmi}{\hbar}\sv\cdot\mathbf{p}}\Big[V\Big(\xv+\frac{\sv}{2}\Big)-V\Big(\xv-\frac{\sv}{2}\Big)\Big]
\label{WP}
\end{equation}
The scalar potential $\phi=V/e$, with the electron charge $e$, and the canonical momentum operator $-\rmi\hbar\nabla$, are fundamental for this picture. 
%What remains beyond the focus is the implicit choice of the gauge where the vector potential $\Av$ is chosen to be zero.  
The choice of the gauge is implicitly assumed, i.e., the vector potential $\Av$ is chosen to be zero.  
However, any other couple $\Av',\phi'$ satisfying $\Av'=\Av+\nabla\chi,\quad\phi'=\phi-\partial\chi/\partial t$ for a given function $\chi$ modifies the Hamiltonian and may lead to a very different physical picture, despite that the electromagnetic environment $\Bv=\nabla\times\Av$, $\Ev=-\nabla\phi-\partial\Av/\partial t$ remains independent on $\chi$~\cite{jackson2002lorenz}. An example is related to  electrons governed by an electric field $\Ev$~\cite{Bloch} in a periodic potential. If Wannier-Stark localized states~\cite{Wannier} are used for the description, the picture involves a discrete energy spectrum accounting for the translational crystal symmetry. If accelerated Bloch states (Houston states)~\cite{Houston} are used, the picture of  continuous acceleration of the wave vector in the crystal band structure gives rise to a periodic electron motion, called Bloch oscillations. % This polarized the  research community speculating on the correctness of the former  or the latter approach.
It has been shown that the two pictures are equivalent and related to the choice of
a vector $(\Av=-\Ev t;~\phi=0)$,
or a scalar potential gauge 
$ (\Av=0, \phi=-\Ev\xv)$, linked by $\chi=-\Ev\xv t$~\cite{krieger1986time, rossi1998bloch}.
%\bibitem{Wannier} G.H.~Wannier, Wave Functions and Effective Hamiltonian for Bloch Electrons in an Electric Field, \emph{Physical Review}, vol.~117, pp.~432-439, \doi{10.1103/PhysRev.117.432}, 1960.
%\bibitem{Houston} W.V.~Houston, Acceleration of Electrons in a Crystal Lattice, \emph{Physical Review}, vol.~57, pp.~184-186, \doi{10.1103/PhysRev.57.184}, 1940.
For the standard Wigner picture, the zero vector potential is a convenient choice, because then the canonical momentum $\pv$ and the kinetic momentum $\Pv$ coincide. This is not true anymore in the case of a magnetic field when $\Pv=\pv-e\Av(\xv)$.
In this case, using the kinetic momentum as a phase space variable offers the advantage that the latter is a physical quantity and thus gauge-invariant~\cite{PhysRevA.33.2913, javanainen1987gauge, aharonov1987phase, levanda1994gauge, varro2003gauge, haas2010fluid}.
Inspired by this fact, Stratonovich~\cite{stratonovich1956gauge}
generalized the Weyl transform to
\begin{equation}
    f_{\rm w}(\Pv,\xv)=\int \frac{\rmd\sv}{(2\pi\hbar)^3}\e^{-\frac{\rmi}{\hbar}\sv\cdot[\Pv+\frac{e}{2}\int_{-1}^1\rmd\tau\Av(\xv+\frac{\sv\tau}{2})]}\rho(\xv+\frac{\sv}{2},\xv-\frac{\sv}{2}).
\end{equation}
Now the transform depends on the vector potential, however, the evolution equation for the Wigner function regarding the position and the kinetic momentum depends only on the electromagnetic field $\Ev$, $\Bv$~\cite{nedjalkov2019wigner}. Thus, the Weyl-Stratonovich transform lifts the gauge dependence, offering more physical transparency to the quantum evolution. In the case $\Av=0$, the Weyl-Stratonovich transform equals the Weyl transform and can thus be seen as an extension. For the sake of convenience, we use $\pv$ instead of $\Pv$ to refer to the kinetic momentum for the remainder of this work.

There are two ways to formulate the evolution equation depending on the physical settings. If the physical system is bounded in space, in a domain enclosed in $(-\Lv/2,\Lv/2)$, where $\Lv$ is called coherence length, the momentum space becomes discrete, involving the integer variable $\mv$:    $\Pv_\mv=\mv\Delta \Pv,\quad \mv\in \Z\times\Z,\quad \Delta \Pv=2\pi\hbar/\Lv $. 
In the limit  $\Lv\to\infty$, called long coherence length limit, the momentum becomes continuous~\cite{nedjalkov2022gauge}. For electromagnetic fields with general spatiotemporal dependence,  both formulations are very challenging from a numerical point of view. A computational experience with the treatment of multidimensional sums and integrals is missing. To gain first experiences, we look for simplified physical conditions to reduce the equation's complexity, allowing in particular for the application of analytical approaches. 
%Helpful for choosing such conditions is the fact that for a homogeneous magnetic field certain integrals vanish, 
The fact that for a homogeneous magnetic field certain integrals vanish is helpful for choosing such conditions,
while the  field appears as the magnetic component of the  Lorentz force in the Liouville operator of the reduced equation. This prompts considering the next term in the Taylor expansion of the magnetic field $\Bv(\xv)$, namely linearly dependent magnetic fields. 
Furthermore, in the case of linear electric fields, they complete the force term in the Liouville operator to a full Lorentz force. We can thus formulate the physical settings under consideration: We consider a transport in a two-dimensional ($2D$) plane with coordinates $\xv=(x,y,0)^T$. A magnetic field $\Bv(y)=(0,0,B_0+B_1y)^T$ points perpendicular to the plane and depends linearly on $y$. The electric field  $\Ev(x,y)=(E_xx,E_yy,0)^T$ accelerates the electron in the plane. %We follow the long coherence length theory. 
The obtained equation using the long coherence length limit~\cite{nedjalkov2022gauge} is given by
\begin{eqnarray}
\fl\left(\frac{\partial}{\partial t}
+
\frac{\pv}{m}\cdot
\frac{\partial}{\partial \xv}
%+e\left(\Ev(\xv)+\frac{\Mv\Delta\pv\times\Bv(y)}{m}\right)
+\Fv%(\pv,\xv)
\cdot\frac{\partial}{\partial \pv}\right)f_{\rm w}\bigl(\pv,\xv\bigr)
=
\frac {B_1\hbar^2}m\frac e{12}
\left(
\frac{\partial^2}{\partial p_y^2}\frac{\partial}{\partial x}
%f_{\rm w}\bigl(\pv,\xv\bigr)
-%\frac {B_1\hbar^2}m\frac e{12}
\frac{\partial}{\partial p_x}\frac{\partial}{\partial p_y}
\frac{\partial}{\partial y}\right)f_{\rm w}\bigl(\pv,\xv\bigr).
\label{linWigEqu}
\end{eqnarray}
We note that the Lorentz force $\Fv=e[\Ev(x,y)+\pv\times\Bv(y)/m]$ in the Liouville operator on the left depends on the electromagnetic field. The operator corresponds to a classical motion over Newtonian trajectories, accelerated by the Lorentz force, linearly dependent on the position coordinates. The term on the right-hand side depends only on the magnetic field gradient $B_1$ and consistently vanishes if $B_1\to 0$. This term is responsible for the  quantum character of the evolution process. Indeed, the structure of \eqn{linWigEqu} resembles the standard Wigner equation. The latter consists of the forceless Liouville operator, whose interplay with the
Wigner potential term gives rise to a fully quantum-coherent evolution. Indeed, the equation is equivalent to the von Neumann equation and in a pure state to the Schr\"odinger equation~\cite{dias2004admissible, tatarskiui1983wigner}. % Prata, Tatarski
However, this term is given by the convolution of the Wigner function with $V_{\rm w}$ in \eqn{WP} and thus depends linearly on $f_{\rm w}$. The corresponding term in \eqn{linWigEqu} introduces high-order mixed derivatives and hence has different numerical aspects.
The numerical experience with  the former equation has matured for more than three decades~\cite{frensley1990boundary, Ferry_and_Shifren, querlioz2006improved, novakovic2011transport, sellier2015introduction, signed_partices}. Furthermore, a peculiarity of phase space formulations of quantum mechanics is the ability to use them for further development of heuristic, physics-based models, associated with quantum phenomena and processes.
Good examples are quantum particle models where particles are provided with additional attributes, such as sign or affinity, while the action of the electric potential is interpreted as scattering or as particle generation~\cite{nedjalkov2011wigner}.
In contrast, alternative quantum theories associate physical quantities and quantum processes with formal mathematical expressions, which offer little physical insight (e.g., operator mechanics). 
%This leads not only to the development of novel computational approaches but offers heuristic models, which are associated with quantum processes, usually described by formal mathematical expressions and physics-based pictures. 

This work provides a numerical analysis of \eqn{linWigEqu} and a particle picture with the corresponding quantum evolution. These quantum particles have a numerical origin, however, they bear the basic properties of the physical models of particles in classical mechanics. The additional particle properties carry the quantum information of the evolution.

% Outline short

In Section~\ref{solutionSection}, an iterative solution to (\ref{linWigEqu}) is presented. The strategy is based on transforming the equation to a Fredholm integral equation, which can be solved by a resolvent expansion.
In Section~\ref{MonteCarloSection}, we derive two different Monte Carlo algorithms for the evaluation of the terms in the resolvent expansion. %: The first one is more formal and evaluates $f_{\rm w}$ pointwise using backward parametrization. The other one uses the more transparent (from a physical point of view) forward parametrization and an exponential distribution of the scattering times. 
In Section~\ref{discussionSection}, the key findings of this work are discussed. %We discuss the maximum simulation time of the algorithms from Section~\ref{MonteCarloSection}.

\section{Iterative solution of the gauge-invariant Wigner equation}
\label{solutionSection}
%A way to solve the gauge-invariant Wigner equation for linear electromagnetic fields is to transform it into a Fredholm integral equation of the second kind and solve it iteratively by using a Liouville–Neumann series. 
% New intro
\begin{comment}
To transform the gauge-invariant Wigner equation into a Fredholm integral equation, the phase space variables are assembled in Newtonian trajectories, representing the evolution of a classical electron under the action of the Lorentz force.
The Liouville operator can then be replaced by a total derivative of time.
The term with the high-order derivatives is handled  with finite difference methods.
Integration over time delivers the desired Fredholm equation. Finally, the terms of the resolvent expansion are determined, representing the Wigner function $f_{\rm w}$ and the averaged physical quantities of the system.
This approach involves a backward parameterization of the trajectory in time. A forward parametrization, corresponding to a  forward evolution, is introduced by the adjoint equation, derived in the same section. In both cases, the high-order derivatives term is interpreted as causing scattering events in the phase space.
\end{comment}
% End New intro
We introduce two new time-dependent functions of the Newtonian trajectory, which replace the phase space variables. We use two parameterizations (backward and forward), which yield different representations of the same solution. This is followed by transforming the gauge-invariant Wigner equation \eqn{linWigEqu} into an integral form, i.e., the Fredholm integral equation, by using a finite difference scheme and a resolvent expansion of the Wigner function. We first present the solution for the backward parameterization and afterward for the forward parameterization. For the latter, we define and solve the adjoint formulation of the Fredholm equation. Finally, both solutions are used to evaluate the expectation value of a physical quantity $A$ iteratively.

% We introduce two new time-dependent functions of the Newtonian trajectory, which replace the phase space variables. They can have two parameterizations (backward and forward), which yield different representations of the solution. Then we can transform the Wigner equation \eqn{linWigEqu} into a Fredholm integral equation. We first present the solutions for the backward parameterization and afterward for the forward parameterization. For the latter, we define and solve the adjoint formulation of the Fredholm equation.

\subsection{Newtonian trajectories with backward and forward parameterization}
The two new time-dependent functions of the Newtonian trajectory %replace the Liouville operator with the total derivative of time. They 
are based on the actual physical behavior of an electron governed by the Lorentz force $\Fv$. The parameterization can be done backward and forward in time.\\

\subsubsection{Backward parameterization}
Consider a particle at a time $t$, the position $\xv$, and the momentum $\pv$ as initial values in a force field $\Fv$. From there, one can determine the position and momentum at an earlier time $t'<t$. They are given by the two integral equations
\begin{eqnarray}
   \xv(t';\pv,\xv,t)&:=&\xv - \int_{t'}^t\frac{\pv(\tau;\pv,\xv,t)}{m}\rmd\tau,\label{Newton1}\\
    \pv(t';\pv,\xv,t)&:=&\pv-\int_{t'}^t\Fv\big(\pv(\tau;\pv,\xv,t), \xv(\tau;\pv,\xv,t)\big)\rmd\tau.
\label{Newton2}
\end{eqnarray}
\subsubsection{Forward parameterization}
In this case, the particle is initialized at $t',\pv',\xv'$. $\pv$ and $\xv$ are then evaluated at a later time $t>t'$ as
\begin{eqnarray}
    &&\xv'(t;\pv',\xv',t'):=\xv'+ \int_{t'}^t\frac{\pv'(\tau;\pv',\xv',t')}{m}\rmd\tau,\\
    &&\pv'(t;\pv',\xv',t'):=\pv'+\int_{t'}^t\Fv\big(\pv'(\tau;\pv',\xv',t'), \xv'(\tau;\pv',\xv',t')\big)\rmd\tau.
\label{Newtonfor1}
\end{eqnarray}
For convenience, we will write $\xv(t'), \pv(t')$ 
 and $\xv'(t), \pv'(t)$ respectively.
We also will use the Liouville theorem, stating that the phase space volume remains constant along the trajectories of the system, i.e., $\int \rmd\pv \rmd\xv=\int \rmd\pv(t') \rmd\xv(t')=\int \rmd\pv'(t) \rmd\xv'(t)$. 
\subsection{Fredholm integral representation of the gauge-invariant Wigner equation}
Next, we show how the gauge-invariant Wigner equation is transformed into an integral form, i.e., the Fredholm integral equation. For this purpose, a finite difference scheme is used to replace the derivatives.
\subsubsection{Integral form}
For the transformation, the variables $\xv$ and $\pv$ in (\ref{linWigEqu}) are replaced by the functions (\ref{Newton1}) and (\ref{Newton2}), respectively. That way, the Liouville operator on the left-hand side can be replaced by a total derivative of time and integrated on $t'$ in the limits $(t_0,t)$. By setting  $t_0=0$ (i.e., the time when the initial condition $f_{{\rm w}_0}$ is known) it is obtained 
\begin{eqnarray}
\fl \eqalign{f_{\rm w}\bigl(\pv,\xv,t\bigr)
=& ~\e^{-\int\limits^t_{0}\gamma(\pv(\tau),\xv(\tau))\rmd\tau} f_{{\rm w}_0}\bigl(\pv(0),\xv(0)\bigr)
+
\int_0^{t} \rmd t'
\e^{-\int\limits^t_{t'}\gamma(\pv(\tau),\xv(\tau))\rmd\tau}\label{intDiffEq}
\\
\fl &\cdot\left[
\frac {B_1\hbar^2}m\frac e{12}
\left(
\frac{\partial^3}{\partial p_y^2\partial x}
-\frac{\partial^3}{\partial p_x\partial p_y\partial y}
\right)
+\gamma(\pv(t'),\xv(t'))
\right]f_{\rm w}\bigl(\pv(t'),\xv(t'),t'\bigr).}
\label{eq:integralForm}
\end{eqnarray}
Here, $\gamma$ is an auxiliary function, which is not presented in the differential form of the equation. Indeed, after taking the derivative with respect to $t_0$, the terms containing $\gamma$  cancel exactly. Later, we show that the introduction of $\gamma$ is convenient from a numerical point of view and also has a physical meaning in the quantum particle model under development.

By taking a closer look at (\ref{eq:integralForm}) we can gain insights into the physical background. The linear coefficient $B_1$ of the magnetic field determines the quantum character of the evolution. Consider the case where $B_1=0$ and $\gamma=0$. The equation then simplifies to $f_{\rm w}\bigl(\pv,\xv,t\bigr)=f_{{\rm w}_0}\bigl(\pv(0),\xv(0)\bigr)$. This means that the Wigner function is constant along the trajectories of the system and one can evaluate $f_{\rm w}$ at any time $t$ by tracing the trajectory back to $t=0$, which is in accordance with Liouville's theorem. Indeed, an initial classical particle density in $\rmd\xv(0)\rmd\pv(0)$ evolves along the trajectories until time $t$ without any change.
\subsubsection{Finite difference scheme}
The integral equation (\ref{intDiffEq}) is not yet of Fredholm type as it contains derivatives of the integrand function $f_{\rm w}$. However, they can be approached by a finite difference scheme, which replaces them with linear combinations of $f_{\rm w}$ defined in adjacent phase space points. Here, we apply a central finite difference scheme. This leads to fifteen terms represented by the  indices $\iv=(i_x,i_y),\jv=(j_x,j_y)$ and coefficients $\alpha_{\iv\jv}$, where $i_x,i_y,j_x,j_y\in\{-1,0,1\}$. We also choose $\gamma$ to be a constant:
\begin{equation}
   \gamma=\gamma(\pv(t'),\xv(t')):=\frac {B_1\hbar^2}m\frac e{96(\Delta P)^2\Delta X}=\rm{constant}
   \label{gamma}
\end{equation}
The convenience of this choice will be discussed below.
With the help of integrals over $\pv$ and $\xv$, and the use of $\delta$ functions the equation obtains a mathematically  formal appearance:
 \begin{eqnarray}
 \fl \eqalign{
 f_{\rm w}(\pv,\xv,t)=f_{\rm i}(\pv,\xv,t)+\int_0^\infty \rmd t'\int \rmd\pv'\int \rmd\xv' \K(\pv,\xv,t,\pv',\xv',t')f_{\rm w}(\pv',\xv',t'),\\
    f_{\rm i}(\pv,\xv,t)=\e^{-t\gamma}f_{{\rm w}_0}\bigl(\pv(0),\xv(0)\bigr),\\
    \K(\pv,\xv,t,\pv',\xv',t')=\rm{~}
    \theta(t-t')\gamma\e^{-(t-t')\gamma}\sum_{\iv,\jv}\alpha_{\iv\jv}\delta(\pv(t')+\iv\Delta P-\pv', \xv(t')+\jv\Delta X-\xv').}\label{eq:fh1}
\end{eqnarray}   
The Heaviside function on time takes care of the proper upper limit $t$.
The detailed form of the kernel $\K$ can be found in \ref{writtenOut}.
\subsection{Solution of the Fredholm integral equation}
In this section, we present a solution for (\ref{eq:fh1}) and how it can be used to evaluate the expectation value of a physical quantity $A$ of a particle. The weak formulation of this task is given as a series of integrals. This series arises from the resolvent expansion of the Wigner function. Consequently, the solution for the physical quantity is done iteratively.

\subsubsection{Weak formulation of the task}
The Wigner function is a quasi-distribution function and can be used as a probability density for quantum particles~\cite{tatarskiui1983wigner}. Consider an arbitrary physical quantity $A$, which depends on position, momentum, and time. The expectation value of $A$ at a time $T$ can be evaluated by
\begin{equation}
    \langle A\rangle(T)=\int_0^\infty \rmd t\int \rmd\pv\int \rmd\xv f_{\rm w}(\pv,\xv,t)A(\pv,\xv,t)\delta(T-t).
    \label{eq:genSolution}
\end{equation}
For convenience reasons we set $A_T(\pv,\xv,t):=A(\pv,\xv,t)\delta(T-t)$.
The solution of Fredholm integral equations is presented by its resolvent expansion~\cite{arfken1999mathematical}, as given in \ref{RE}.
It allows to represent   $\langle A\rangle(T)$ as a series
\eq
    \eqalign{\fl\langle A\rangle(T)=\sum_{n=0}^\infty\int_0^\infty \rmd t\int \rmd\pv\int \rmd\xv f_n(\pv,\xv,t)A_T(\pv,\xv,t)
    =\sum_{n=0}^\infty\langle A\rangle_n(T).}
    \label{eq:LiouvilleSeries}
\eqend
In particular, if $A$ is chosen to be a delta function, the series yields the expansion of the Wigner function.
\subsubsection{Resolvent expansion of the Wigner function}
%Before we present the solution of the terms $\langle A\rangle_n(T)$, we introduce two new functions to make it more comprehensible. Afterward, these functions are used in the solution of the Wigner function. The consecutive application of the kernel onto itself causes the trajectory to scatter several times, i.e.,$\pv$ and $\xv$ make discontinuous jumps. 
Given the scattering indices $(\iv_k,\jv_k)_{1\le k\le n}$ and the scattering times $t_1<t_2<\dots<t_n$, we introduce the trajectory with scattering events for backward parameterization as
\begin{figure}[b]
    \begin{center}
    \includegraphics[width=0.7\columnwidth]{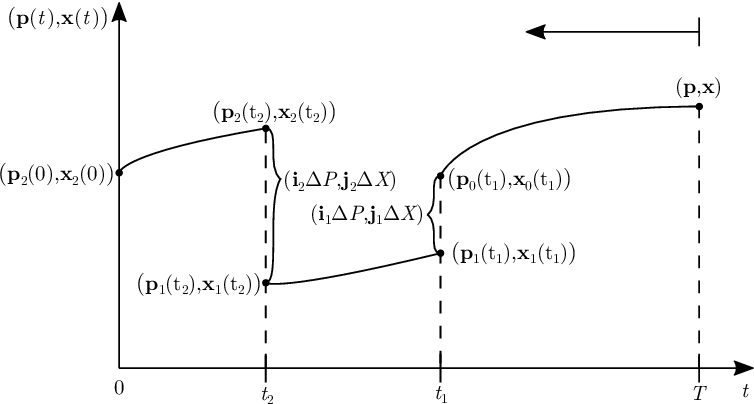}   
    \end{center}
    \caption[Trajectory of the 2nd iteration with backward parameterization]
    {Trajectory of the 2nd iteration with backward parameterization
    }
    \label{fig:trajectoryBW}
\end{figure}
\begin{eqnarray}
    \eqalign{
    \fl \pv_n\big(t'\big):=\cases{\pv_{n-1}(t') & for $t_n<t'\le T$\\
\pv\big(t';\pv_{n-1}(t_{n})+\iv_n\Delta P,\xv_{n-1}(t_{n})+\jv_n\Delta X,t_n\big)&for $0\le t'\le t_n$,\\}\\
    \fl \xv_n\big(t'\big):=\cases{\xv_{n-1}(t') & for $t_n<t'\le T$\\
\xv\big(t';\xv_{n-1}(t_{n})+\iv_n\Delta P,\xv_{n-1}(t_{n})+\jv_n\Delta X,t_n\big)&for $0\le t'\le t_n$,\\}
}
    \label{eq:XnPnBw}
\end{eqnarray}
where we use the convention 
$\pv_0(t'):=\pv(t';\pv,\xv,T)$, $\xv_0(t'):=\xv(t';\pv,\xv,T)$.

%A depiction of these functions can be seen in figure~\ref{fig:trajectoryBW}.
\begin{comment}
Now we use the trajectories to derive the iterative solution of the Wigner function. By applying the general solution (\ref{eq:liouvilleNeumann1}) of the Fredholm equation to our problem, we get
\begin{eqnarray}
    \eqalign{\flf_0(\pv,\xv,t)=&f_{\rm i}(\pv,\xv,t),\\
    \flf_n(\pv,\xv,t)=&\int_0^\infty dt_1\int d\pv_1\int d\xv_1\dots\int_0^\infty dt_{n}\int d\pv_{n}\int d\xv_{n}\K(\pv,\xv,t,\pv_{1},\xv_{1},t_{1})\\
    \fl&\cdot\prod_{k=1}^{n-1}\K(\pv_{k},\xv_{k},t_{k},\pv_{k+1},\xv_{k+1},t_{k+1})f_{\rm i}(\pv_{n},\xv_{n},t_{n})}
    \label{eq:bwsolution}
\end{eqnarray}
\end{comment}
In accordance with  (\ref{eq:liouvilleNeumann1}) we obtain
\eq
    \eqalign{f_n(\pv,\xv,t)=&\gamma^n\e^{-\gamma t}\int_0^t\rmd t_1\int_0^{t_1}\rmd t_2\dots\int_0^{t_{n-1}}\rmd t_n\\&\sum_{\iv_1,\jv_1}\dots\sum_{\iv_n,\jv_n}\prod_{k=1}^n(\alpha_{\iv_k,\jv_k})f_{{\rm w}_0}\big(\pv_n(0),\xv_n(0)\big),}
    \label{eq:fn}
\eqend
where $f_0(\pv,\xv,t)=\e^{-t\gamma}f_{{\rm w}_0}\big(\pv(0),\xv(0)\big)$, see \ref{prooff_w}.
%where the proof can be found in \ref{prooff_{\rm w}}.

The existence of the backward Newtonian trajectories
invokes a picture of a pointlike particle that evolves back in time. The delta functions, which give rise to offsets of the phase space positions, can be interpreted as scattering factors. Figure~\ref{fig:trajectoryBW} schematically presents the second term in the iterative expansion of the Wigner function.
%The trajectory of $\pv(t)$ and $\xv(t)$ is projected onto the $y$-Axis. 
The particle starts at $(\pv,\xv,T)$ and moves back in time in the phase space according to the Lorentz force $\Fv$. When the particle reaches $t_1$ it is scattered, i.e., a factor $(\iv_1\Delta P,\jv_1\Delta X)$ is added. Next, it follows the trajectory again until it reaches $t_2$. This process is repeated until $t=0$ is reached.
\subsubsection{Iterative representation of physical quantities}
To evaluate the solution of $\langle A\rangle_n(T)$, we insert the solution of $f_n,n\in\N$ in (\ref{eq:fn}) into (\ref{eq:LiouvilleSeries}). This yields
\begin{eqnarray}
 \eqalign{\fl   \langle A\rangle_n(T)
=\gamma^n\e^{-T\gamma}\int \rmd\pv\int \rmd\xv \int_0^T \rmd t_1\int_0^{t_1} \rmd t_2\dots\int_0^{t_{n-1}} \rmd t_nA(\pv,\xv,T)\\ 
\cdot\sum_{\iv_1,\jv_1}\dots\sum_{\iv_n,\jv_n}\prod_{k=1}^n(\alpha_{\iv_k\jv_k})f_{{\rm w}_0}\big(\pv_n(0),\xv_n(0)\big).}\label{An}
\end{eqnarray}
This shows us how each element $\langle A\rangle_n(T)$ is generated. 
In the backward parameterization case, the trajectory of $\pv$ and $\xv$ starts at $T$ and goes back in time, according to (\ref{eq:XnPnBw}). The particle is scattered at each $(t_i)_{i\in\{1,2,\dots,n\}}$, where $T>t_1>t_2>\dots>t_n>0$. The indices $\iv_k$ and $\jv_k$  are implicitly included in the functions $\pv_n$ and $\xv_n$. Reaching the final momentum and position at $t=0$, they are used as the arguments of the initial condition of the Wigner function $f_{{\rm w}_0}$. The integration limits of the $t_i$'s and consequently their orders are determined by the $\theta$ functions of the kernel.
% In the backward parameterization case, the trajectory of $\pv$ and $\xv$ starts at $T$ and goes back in time, according to (\ref{Newton1}), (\ref{Newton2}). Then, the particle is scattered at each $(t_i)_{i\in\{1,2,\dots,n\}}$, where $T>t_1>t_2>\dots>t_n>0$. The integration limits of the $t_i$'s and consequently their orders are determined by the $\theta$-functions of the kernel. At each scattering event, the trajectory splits up, and the terms $\iv_k\Delta P$ and $\jv_k\Delta X$ are added, which are implicitly included in the functions $\pv_n$ and $\xv_n$. Afterward, the trajectory of the new point is followed again until the next $t_i$. This process is repeated until the initial time $0$ is reached, where the initial condition of the Wigner function $f_{{\rm w}_0}$ is evaluated at the final point. 

% The same picture is associated with the expansion of a  generic physical quantity.

%If one is interested in the solution of the Wigner function $f_{\rm w}$ at a point $\pv_0,\xv_0,T$, $A$ can be set to $A(\pv,\xv,t):=\delta(\pv_0-\pv)\delta(\xv_0-\xv)$.

\subsection{Solution of the adjoint integral equation}
In this section, a solution of the Fredholm integral equation (\ref{eq:fh1}) is presented where forward parameterization is used. The weak formulation of this task is given by the adjoint formulation of the Fredholm integral equation. Finally, the solution for the adjoint equation is used to derive the expectation value of a physical quantity iteratively.
\subsubsection{Weak formulation of the task}

The adjoint of a Fredholm integral equation has the same kernel, but the
integration is over the other set of variables:
\begin{equation}
    \fl g(\pv',\xv',t')=g_{\rm i}(\pv',\xv,'t')+\int_{0}^\infty \rmd t\int_{-\infty}^\infty \rmd\pv\int_{-\infty}^\infty \rmd\xv\K(\pv,\xv,t,\pv',\xv',t')g(\pv,\xv,t)\label{adjointProblem1}
\end{equation}
The free term $g_{\rm i}$ can be determined from the weak formulation of the task, namely to find the expecation value of a physical quantity $A$. 
The following relation follows from the exchange Lemma in \ref{adjointLemma} and the Liouville theorem. By choosing $g_{\rm i}(\pv',\xv',t'):=A_T(\pv',\xv',t')$ we can show
\begin{eqnarray}
\eqalign{
&\fl{\langle A\rangle(T)=\int\limits_0^\infty \rmd t\int \rmd\pv\int \rmd\xv f_{\rm w}(\pv,\xv,t)A_T(\pv,\xv,t)
    =\int\limits_0^\infty \rmd t\int \rmd\pv\int \rmd\xv f_{\rm w}(\pv,\xv,t)g_{\rm i}(\pv,\xv,t)}
    \\
    &\fl =\int\limits_0^\infty \rmd t\int \rmd\pv\int \rmd\xv f_{\rm i}(\pv,\xv,t)g(\pv,\xv,t)
    =\int\limits_0^\infty \rmd t\int \rmd\pv\int \rmd\xv \e^{-t\gamma}f_{{\rm w}_0}\big(\pv,\xv\big)g(\pv'(t),\xv'(t),t).}
\end{eqnarray}
Like before, we consider the resolvent expansion to evaluate $\langle A\rangle(T)$, which yields
\begin{equation}
\fl
    \langle A\rangle(T)=\sum_{n=0}^\infty\int_0^\infty \rmd t\int \rmd\pv\int \rmd\xv \e^{-t\gamma}f_{{\rm w}_0}(\pv,\xv)g_n(\pv'(t),\xv'(t),t)
    =\sum_{n=0}^\infty\langle A\rangle_n(T).
    \label{eq:ATfw}
\end{equation}
The integration over the other set of variables $\pv,\xv,t$ gives rise to a transition to a forward parametrization of the arguments in the  $\delta$ functions in the kernel:
\begin{eqnarray}
\eqalign{
    \fl\delta\big(\pv(t')+\iv\Delta P-\pv',\xv(t')+\jv\Delta X-x'\big)\\
    % \fl=\delta\big(\pv-\int_{t'}^t\Fv(\pv(\tau), \xv(\tau))d\tau+\iv\Delta P-\pv',\xv - \int_{t'}^t\frac{\pv(\tau)}{m}d\tau+\jv\Delta X-\xv'\big)\\
    \fl\hskip 1cm =\delta\big(\pv-\pv'(t;\pv'-\iv\Delta P,\xv'-\jv\Delta X,t'),\xv-\xv'(t;\pv'-\iv\Delta P,\xv'-\jv\Delta X,t')\big)}
\end{eqnarray}
\begin{comment}
where backward parameterization was used on the left side and forward parameterization on the right. Here we have used the fact that % Explanation: trajectory + uniqueness of solution
\begin{eqnarray*}
    \fl\pv(t')=\pv'-\iv\Delta P{\rm\quad and\quad}\xv(t')=\xv'-\jv\Delta X\\
    \fl\Longleftrightarrow\pv'(t,\pv(t'),\xv(t'),t')=\pv'(t;\pv'-\iv\Delta P,\xv'-\jv\Delta X,t')\\
    \fl{\rm and~~}
    \xv'(t,\pv(t'),\xv(t'),t')=\xv'(t;\pv'-\iv\Delta P,\xv'-\jv\Delta X,t')\\
    \fl\Longleftrightarrow\pv=\pv'(t;\pv'-\iv\Delta P,\xv'-\jv\Delta X,t')
    {\rm\quad and\quad}
    \xv=\xv'(t;\pv'-\iv\Delta P,\xv'-\jv\Delta X,t').
\end{eqnarray*}
The kernel is still the same, but it is differently formulated. 
% Before, we followed the trajectory, starting at $(\pv,\xv,t)$ backward in time, then scattered it and compared it with $(\pv',\xv')$. This time, starting from $(\pv',\xv',t')$, the scattering takes place first, and then the trajectory is followed forward in time until $t$, where it is compared to $(\pv,\xv)$. 
\end{comment}
For $\K$ this yields
\begin{eqnarray}
\eqalign{
    \fl\K(\pv,\xv,t,\pv',\xv',t')=\theta(t-t')\gamma\e^{-(t-t')\gamma}\sum_{\iv,\jv}\alpha_{\iv\jv}\\ \fl\hskip 1cm \cdot\delta\big(\pv-\pv'(t;\pv'-\iv\Delta P,\xv'-\jv\Delta X,t'),\xv-\xv'(t;\pv'-\iv\Delta P,\xv'-\jv\Delta X,t')\big).}\label{eq:fh4}
\end{eqnarray}
\subsubsection{Solution for the adjoint equation }
\begin{figure}[h]
    \begin{center}
    \includegraphics[width=0.7\columnwidth]{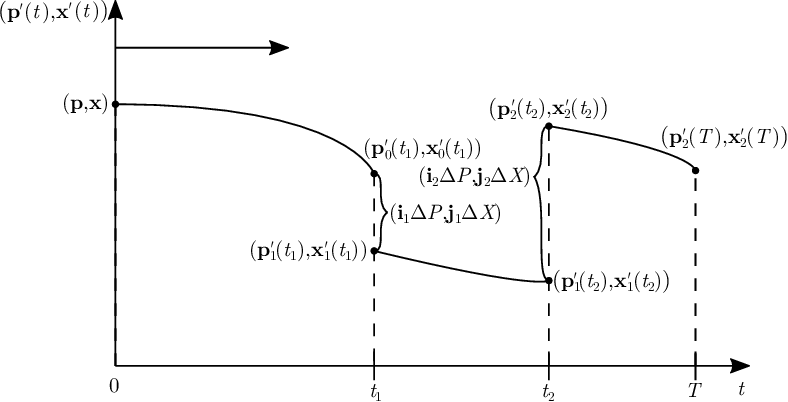}
    \end{center}
    \caption[Trajectory of the 2nd iteration with forward parameterization]{Trajectory of the 2nd iteration with forward parameterization. %\par \small 
    %The trajectory of $\pv(t)$ and $\xv(t)$ is projected onto the $y$-Axis. 
%    The particle starts at $(\pv,\xv,0)$ and moves forward in time in the phase space according to the Lorentz force $\Fv$. When the particle reaches $t_1$ it is scattered, i.e., a factor $(\iv_n\Delta P,\jv_n\Delta X)$ is subtracted. Afterward, it follows the trajectory again until it reaches $t_2$. This process is repeated until $t=T$ is reached, where $A_T$ is evaluated.
}
    \label{fig:trajectoryFW}
\end{figure}

We introduce the trajectory with scattering events for  forward parameterization.
Given the scattering indices $(\iv_k,\jv_k)_{1\le k\le n}$ and the scattering times $t_1<t_2<\dots<t_n$, we use the convention
 $\pv_0'(t):=\pv(t;\pv,\xv,0)$, $\xv_0'(t):=\xv(t;\pv,\xv,0)$
 to define
\begin{eqnarray}
    \eqalign{
    \fl \pv_n'\big(t\big):=\cases{\pv_{n-1}'(t) & for $0\le t \le t_n$\\
\pv'\big(t;\pv'_{n-1}(t_{n})-\iv_n\Delta P,\xv'_{n-1}(t_{n})-\jv_n\Delta X,t_n\big)&for $t_n < t\le T$,\\}\\
    \fl \xv_n'\big(t\big):=\cases{\xv_{n-1}'(t) & for $0\le t \le t_n$\\
\xv'\big(t;\pv'_{n-1}(t_{n})-\iv_n\Delta P,\xv'_{n-1}(t_{n})-\jv_n\Delta X,t_n\big)&for $t_n < t\le T$.\\}
}
\end{eqnarray}
A depiction of these functions can be seen in Figure~\ref{fig:trajectoryFW}. 
The resolvent series for the solution is then presented by the term
\begin{comment}
As with the solution of the Wigner function, we use the new trajectories we have just defined for the adjoint equation function $g$. By applying the general solution (\ref{eq:liouvilleNeumann4}) of the Fredholm equation to our problem, we get
\begin{eqnarray}
\eqalign{
    \flg_0(\pv,\xv,t)=&g_{\rm i}(\pv,\xv,t),\\
    \flg_n(\pv,\xv,t)=&\int_0^\infty dt_1\int d\pv_1\int d\xv_1\dots\int_0^\infty dt_{n}\int d\pv_{n}\int d\xv_{n}\K(\pv,\xv,t,\pv_{n},\xv_{n},t_{n})\\ &\cdot\prod_{k=1}^{n-1}\K(\pv_{k+1},\xv_{k+1},t_{k+1},\pv_{k},\xv_{k},t_{k})g_{\rm i}(\pv_1,\xv_1,t_1).}
    \label{eq:gsolution}
\end{eqnarray}
For reasons of notation, the order of the time variables is changed in the following. $t$ becomes the innermost variable of the integrals, $t_1$ becomes the time parameter of $g_n$ and all other time variables $t_i,i\in{1,2,\dots,n}$ are shifted to the left. Inserting $A_T$ and (\ref{eq:fh4}) into $g_{\rm i}$ and $\K$ respectively yields
\end{comment}
\begin{eqnarray}
    \eqalign{g_n(\pv'(t_1),\xv'(t_1),t_1)=&~\gamma^n\e^{-(T-t_1)\gamma}\int\limits_{t_1}^T \rmd t_2\dots\int\limits_{t_{n-1}}^T \rmd t_n\\ &\sum_{\iv_1,\jv_1\ldots\iv_n,\jv_n
    }
    %\dots\sum_{\iv_n,\jv_n}
\prod_{k=1}^n(\alpha_{\iv_k,\jv_k})A_T\big(\pv_n(T),\xv_n(T),T\big),}
    \label{eq:gn}
\end{eqnarray}
with $g_0(\pv'(t),\xv(t),t)=A_T(\pv_0(t),\xv_0(t),t)$.
%which can be shown analogously to the proof for the solution of $f_{\rm w}$ in \ref{prooff_{\rm w}}. 
\subsubsection{Iterative representation of physical quantities}
The series for the expectation value of a physical quantity is obtained by inserting (\ref{eq:gn})  into (\ref{eq:ATfw}).
The general term is then
\begin{comment}
\eqalign{
    \fl\langle A\rangle_0(T)&=\int_0^\infty \rmd t\int \rmd\pv\int \rmd\xv \e^{-t\gamma}f_{{\rm w}_0}\big(\pv,\xv\big) g_0(\pv'(t),\xv'(t),t)\\
    \fl&=\int_0^\infty \rmd t\int \rmd\pv\int \rmd\xv \e^{-t\gamma}f_{{\rm w}_0}\big(\pv,\xv\big) A_T(\pv_0'(t),\xv_0'(t),t)\\
    \fl&=\e^{-T\gamma}\int \rmd\pv\int \rmd\xv f_{{\rm w}_0}\bigl(\pv,\xv\bigr)A(\pv_0'(T),\xv_0'(T),T),\\
    \fl\langle A\rangle_n(T)&=\int_0^\infty dt_1\int \rmd\pv\int \rmd\xv \e^{-t_1\gamma}f_{{\rm w}_0}(\pv,\xv)g_n(\pv'(t_1),\xv'(t_1),t_1)\\
    \fl&=\int_0^\infty dt_1\int \rmd\pv\int \rmd\xv \e^{-t_1\gamma}f_{{\rm w}_0}(\pv,\xv)\gamma^n\e^{-(T-t_1)\gamma}\\ 
    \fl&\cdot\int_{t_1}^Tdt_2\int_{t_2}^Tdt_3\dots\int_{t_{n}}^Tdt\sum_{\iv_1,\jv_1}\sum_{\iv_2,\jv_2}\dots\sum_{\iv_n,\jv_n}\prod_{k=1}^n(\alpha_{\iv_k,\jv_k})A_T\big(\pv_n'(T),\xv_n'(T)\big)\\
    \fl&=\gamma^n\e^{-T\gamma}\int \rmd\pv\int \rmd\xv f_{{\rm w}_0}(\pv,\xv)\\
    \fl&\cdot\int_0^Tdt_1\int_{t_1}^Tdt_2\dots\int_{t_{n-1}}^Tdt_n\sum_{\iv_1,\jv_1}\sum_{\iv_2,\jv_2}\dots\sum_{\iv_n,\jv_n}\prod_{k=1}^n(\alpha_{\iv_k\jv_k})A\big(\pv_n'(T),\xv_n'(T)\big).} \label{Anfw}
\end{comment}
\begin{eqnarray}
\eqalign{\langle A\rangle_n(T)
=&\gamma^n\e^{-T\gamma}\!
\int\! \rmd\pv\!\int \!\rmd\xv f_{{\rm w}_0}(\pv,\xv)\!
    \int\limits_0^T\!\rmd t_1\dots\!\int\limits_{t_{n-1}}^T\!\rmd t_n\\ &\sum_{\iv_1,\jv_1\ldots\iv_n,\jv_n}\prod_{k=1}^n(\alpha_{\iv_k\jv_k})A\big(\pv_n'(T),\xv_n'(T),T\big).}
    \label{Anfw}  
\end{eqnarray}
Since both (\ref{Anfw}) and (\ref{An}) are transformations of the general solution (\ref{eq:genSolution}), they are indeed equivalent. Equation (\ref{Anfw}) remarkably resembles the corresponding expression for the Monte Carlo averages of an ensemble of $M$ classical (Boltzmann) electrons, which move under the action of the Lorentz force and are scattered by, e.g., lattice vibrations (phonons)~\cite{bookStochastic}. They are point-like particles with an initial distribution $f_{{\rm w}_0}$, which initializes the starting phase space points $\pv,\xv$. They determine Newtonian trajectories followed % by the accelerated 
by the force particles during their free flight. The free flight is interrupted by scattering events, which, at a time $t_1$, update the phase-space coordinates.
The latter initialize a novel piece of Newtonian trajectory 
for the next free flight. The evolution continues until the time $T$ is reached and then each particle $l$ contributes with its current value  $A_l$ (e.g., velocity, energy) to the statistical sum $\sum_l^MA_l$, which evaluates $\langle A\rangle$.
The process corresponds to the scheme depicted in Figure~\ref{fig:trajectoryFW}, which suggests a picture where pointlike quantum particles follow the same sequence of events. However, several problems need to be addressed to associate (\ref{Anfw}) with a quantum particle model. The classical initial distribution is non-negative, $f_{{\rm w}_0}\ge 0$, while in the quantum case, $f_{{\rm w}_0}$ could be any legitimate Wigner function and thus allows for negative values. This affects the evaluation of the physical averages, as can be seen already from the zeroth order term, which dominates if the evolution time is much smaller than the mean scattering time: In order to account for the sign, the statistical sum for the envisaged quantum particle model  must be generalized to  $\sum_l^Mw_lA_l$
where the quantity $w_l$, called weight, should carry the sign of $f_{{\rm w}_0}$ in the point of initialization of the $l$-th particle. Next, in the classical evolution, the scattering time (e.g., $t_1$) exponentially depends on the frequency of interaction with phonons, while in the quantum counterpart the sequence $t_1<t_2<\cdots$ is predetermined. This suggests looking for an analogical physical interpretation of the prefactor in (\ref{Anfw}). Finally, both classical and quantum counterparts rely on Newtonian trajectories, and hence the difference between the two kinds of evolution is due to the scattering: A fundamental difference between classical and quantum scattering is expected. These problems, formulated by heuristic considerations, are rigorously addressed next by the rules of the Monte Carlo theory for integration. %two  Monte Carlo algorithms.

%The procedure is similar to (\ref{An}). Only this time, the trajectory is followed forward in time, with $0<t_1<t_2<...<t_n<T$ based on the $\theta$-functions of the kernel. Again, the particle is scattered several times, where the terms $\iv \Delta P$ and $\jv \Delta X$ are subtracted, implicitly included in $\pv_n'$ and $\xv_n'$. This process is repeated until the particle reaches its final point at $T$, where $A$ is evaluated instead of $f_{\rm i}$.
\section{Monte Carlo algorithms}
\label{MonteCarloSection}
% New intro
%In this section, we derive two different Monte Carlo algorithms for the evaluation of the terms in the resolvent expansion: 
The two algorithms presented in this section differ in both, their parameterization and the distribution of the scattering times.
The first one is more formal and evaluates $f_{\rm w}$ pointwise using backward parametrization and a uniform distribution for the scattering times.  The other one uses the more transparent (from a physical point of view) forward parametrization and introduces an exponential distribution of the scattering, which is a characteristic of the evolution of classical particles in the presence of scattering events.
This gives rise to a quantum particle model, where the evolution of pointlike particles consists of consecutive events of free-flight over the Lorentz force-governed Newtonian trajectories, followed by scattering events. 
% An important quantity is the weight of the particles corresponding to their share in the statistical evaluation of the physical averages.  The weight of the classical particles is always 1. Quantum particles change both, their weight and their sign. This is a property of quantum evolution in the presence of electric potentials. Their weight and sign are changed by the processes of scattering associated with the Wigner potential.
% A peculiarity of these cases is that scattering is local in space and leads to a change of momentum. 
% In contrast, in the presence of a magnetic field, we show that the term introduces spatial transitions, giving rise to scattering associated with the magnetic field being nonlocal in space.
% End New Intro
\subsection{Backward algorithm}%{Backward parameterization}
\begin{figure}[h]
\begin{center}
\begin{tikzpicture}[scale=0.55, transform shape]
\node (start) [startstop2] {Start};
\node (init) [process2, below=0.5cm of start] 
{Initialization: $N,(N_n)_{n\in\{0,1,2,\dots,N\}};n\leftarrow 0,$\\
$(A_n)_{n\in\{0,1,2,\dots,N\}}\leftarrow\vec{0},j\leftarrow 1,$\\$t_0\leftarrow T$
};
\node (step1) [process2, below=0.5cm of init] 
{$i\leftarrow 0$};
\node (nNeq0) [decision2, below=0.5cm of step1] {$n= 0$};
\node (time1) [process2, below=0.5cm of nNeq0] {$i\leftarrow i+1$,\\$t_i\sim U(0,t_{i-1})$};
\node (iEq0) [decision2, below=0.5cm of time1] {$i<n$};
\node (step2) [process2, below=0.5cm of iEq0] {$s\leftarrow 1,i\leftarrow 0$\\$\pv,\xv\sim P(\pv,\xv)$\\$\pv_T\leftarrow\pv,\xv_T\leftarrow\xv$};
\node (step3) [process2, right=5cm of start] {$\iv,\jv\sim P_{\iv,\jv}$,\\$s\leftarrow s\cdot \gamma t_{i-1}|\alpha|{\rm sign}(\alpha_{\iv\jv})$\\$\pv\leftarrow \pv+\iv\Delta P,$\\$\xv\leftarrow\xv+\jv\Delta X$};
\node (endPos) [decision2, below=0.5cm of step3] {$i=n$};
\node (step4) [process2, right=1cm of endPos] {$\pv\leftarrow\pv(t_{i+1};\pv,\xv,t_i),$\\$\xv\leftarrow\xv(t_{i+1};\pv,\xv,t_i),i\leftarrow i+1$};
\node (step5) [process2, below=0.5cm of endPos] {$\pv\leftarrow\pv(0;\pv,\xv,t_n),$\\$\xv\leftarrow\xv(0;\pv,\xv,t_n)$};
\node (step6) [process2, below=0.5cm of step5] {$A_n\leftarrow A_n+s\e^{-T\gamma} f_{{\rm w}_0}(\pv,\xv)\frac{A(\pv_T,\xv_T,T)}{P(\pv_T,\xv_T)}$};
\node (jEqNn) [decision2, below=0.5cm of step6] {$j<N_n$};
\node (step7) [process2, right=1cm of jEqNn] {$j\leftarrow j+1$};
\node (nEqN) [decision2, below=0.5cm of jEqNn] {$n=N$};
\node (step8) [process2, right=1cm of nEqN] {$j\leftarrow 1,n\leftarrow n+1$};
\node (stop) [startstop2, below=0.5cm of nEqN] {Return $\leftarrow\sum_{n=0}^N\frac{A_n}{N_n}$};

\draw [->] (start) -- (init);
\draw [->] (init) -- (step1);
\draw [->] (step1) -- (nNeq0);
\draw [->] (nNeq0) -- node[anchor=west] {no} (time1);
\draw [-*] (nNeq0.east) -- node[anchor=south] {yes} +(3.25cm,0);
\draw [->] (time1) -- (iEq0);
\draw [->] (iEq0) -- node[anchor=west] {no} (step2);
\draw [->] (iEq0.east) -- node[anchor=south] {yes} +(2cm,0) |- (time1);
\draw [->] (step2.east) -- +(2cm,0) |- (endPos);
\draw [->] (step3) -- (endPos);
\draw [->] (endPos) -- node[anchor=south] {no} (step4);
\draw [->] (step4.north) |- (step3.east);
\draw [->] (endPos) -- node[anchor=west] {yes} (step5);
\draw [->] (step5) -- (step6);
\draw [->] (step6) -- (jEqNn);
\draw [->] (jEqNn) -- node[anchor=south] {yes} (step7);
\draw [-*] (step7.east) -| +(1,-2.45cm);
\draw [->] (jEqNn) -- node[anchor=west] {no} (nEqN);
\draw [->] (nEqN) -- node[anchor=south] {no} (step8);
\draw [->] (step8.east) -| +(1cm,-3.5cm) -- +(-21cm,-3.5cm) |- (step1.west);
\draw [->] (nEqN) -- node[anchor=west] {yes} (stop);
\end{tikzpicture}
\end{center}
\caption{Flow chart of the backward algorithm}
\label{fig:flowchartBW}
\end{figure}
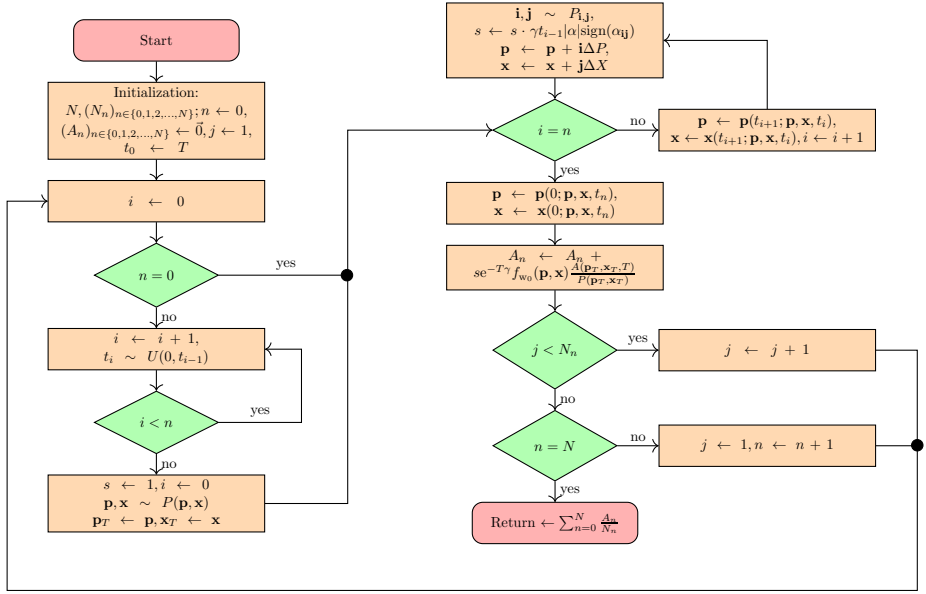

The Monte Carlo algorithm introduced in this section allows to evaluate the terms $\langle A\rangle_n(T)$ of the resolvent expansion in (\ref{An}).
%In this section, the Monte Carlo method is used to evaluate the terms $\langle A\rangle_n(T)$ of the resolvent expansion for backward parameterization in (\ref{An}), and we present an algorithm based on it. 
For this purpose, the integrals and sums are expressed as an expectation value $E[X_n]$ with a probability density $P_{X_n}$ and a random variable $X_n$.
% \subsubsection{Distribution of the random points}
\label{distribution}
The terms $\langle A\rangle_n(T)$ are set to
\begin{eqnarray}
    \eqalign{\langle A\rangle_n(T)=E[X_n]=&\int \rmd\pv\int \rmd\xv \int_0^T \rmd t_1\int_0^{t_1} \rmd t_2\dots\int_0^{t_{n-1}} \rmd t_n \\&\sum_{\iv_1,\jv_1}\sum_{\iv_2,\jv_2}\dots\sum_{\iv_n,\jv_n}P_{X_n}X_n.}
    \label{expVal1}
\end{eqnarray}
$P_{X_n}$ acts as a selector for the scattering indices $(\iv_k,\jv_k)_{1\le k\le n}$, the scattering times $t_1<t_2<\dots<t_n$, and the initial points $\pv,\xv$ of the trajectory. Thus, it is split into a product of three probability functions:
\begin{itemize}
    \item For the coefficients $\alpha_{\iv\jv}$ of the kernel (\ref{eq:fh1}), we introduce a discrete transition probability $P_{\iv\jv}:=\frac{|\alpha_{\iv\jv}|}{|\alpha|}$, where $|\alpha|:=\sum_{\iv\jv}|\alpha_{\iv\jv}|=41$, see (\ref{eq:kernel}). This means that the direction in which the trajectory scatters is chosen randomly, distributed proportionally to $|\alpha_{\iv\jv}|$. 
    \item For the initial points $\pv,\xv$ of the trajectory, a density function $P$ is chosen. Both $A$ and $f_{{\rm w}_0}$ depend on $\pv$ and $\xv$, thus a possible choice could be $P(\pv,\xv)\propto |A(\pv,\xv)f_{{\rm w}_0}(\pv,\xv)|$.
    % Ideally, $P$ should be chosen proportionally to $|A|$ if the integral of $|A|$ over the momentum and position space is known. Otherwise, it should be a good approximation. 
    \item The scattering times $t_1,\dots,t_n$ are evenly distributed on the intervals $(0,T)$ for $t_1$ and on $(0,t_{i-1})$ for $t_i,i\in\{2,\dots,n\}$. The density function of a uniform distribution is normalized by the inverse of the length of the integral, which has to be considered in $X_n$ by the product $T\prod_{i=1}^{n-1}t_i$.
\end{itemize}
% The ideal density function would be
% \begin{equation}
%     P_{\rm{ideal}}(\pv,\xv)=\frac{|A(\pv,\xv,T)f_{{\rm w}_0}\big(\pv_n(0),\xv_n(0)\big)|}{\int \rmd\pv\int \rmd\xv |A(\pv,\xv,T)f_{{\rm w}_0}\big(\pv_n(0),\xv_n(0)\big)|}.
% \end{equation}
% Since we do not know where the trajectory terminates at time $0$ due to the scattering events, 
In combination they yield $P_{X_n}=|\alpha_{\iv\jv}|/|\alpha|P(\pv,\xv)(T\prod_{i=1}^{n-1}t_i)^{-1}$. Since the corresponding random variable $X_n$ is the estimator of $\langle A\rangle_n(T)$, it is evaluated and averaged for several arguments randomly selected according to $P_{X_n}$. To satisfy (\ref{expVal1}) it is given as
\begin{eqnarray}
    \fl X_n=\gamma^n|\alpha|^n\e^{-T\gamma}\frac{A(\pv,\xv,T)}{P(\pv,\xv)} T\prod_{i=1}^{n-1}(t_i)\prod_{k=1}^n\big({\rm sign}(\alpha_{\iv_k\jv_k})\big)f_{{\rm w}_0}\big(\pv_n(0),\xv_n(0)\big).
    \label{AnBwAlgo}
\end{eqnarray}
%\subsubsection{Backward algorithm for a physical quantity}
%Now we are going to develop an algorithm, based on the Monte Carlo method with the random variable and probability functions of the previous segment.\\
The expectation value of a physical quantity can be obtained by Algorithm~\ref{algorithm:algoBW} (see also Figure~\ref{fig:flowchartBW}).
%Algorithm~\ref{algorithm:algoBW} outlines the obtained backward algorithm for a physical quantity. 
\begin{algorithm}{Backward algorithm}%{Backward trajectory algorithm for a physical quantity}
\label{algorithm:algoBW}
\begin{enumerate}[label=\arabic*]
    \item Initialization of $N$, $(N_n)_{n\in\{0,1,2,\dots,N\}},n\leftarrow0$ and a variable, say $(A_n)_{n\in\{0,1,2,\dots,N\}}\leftarrow\vec{0}$. $N$ sets the total number of terms in the iterative expansion (\ref{eq:LiouvilleSeries}). $N_n$ determines the number of independent numerical trajectories with $n$ scattering events. $j\leftarrow 1$ is a counter for $N_n$. $A_n$ represents the value of the $n$-th term in the resolvent expansion. $t_0$ is initialized by $t_0\leftarrow T$.
    \item If $n\neq 0,$ the scattering times $(t_i)_{i\in\{1,2,\dots,n\}}$ are chosen in order because the upper limit of every $t_i$ depends on $t_{i-1}$. Each $t_i\sim {\rm U}(0,t_{i-1})$ is generated randomly,  with the uniform distribution ${\rm U}$ on the interval $(0,t_{i-1})$. $s\leftarrow1$, which represents all factors in $A_n$ that are updated at each scattering event, and $i\leftarrow0$. $(\pv,\xv)\sim P(\pv,\xv)$ are chosen randomly, and distributed according to the chosen probability function $P(\pv,\xv)$. The initial values $\pv_T\leftarrow\pv$ and $\xv_T\leftarrow\xv$ are stored separately. If $n=0$, jump to step 5.
    \item Starting from the current $\pv$ and $\xv$ the trajectory is followed until it reaches the next scattering event at $t_{i+1}$, i.e., $\pv\leftarrow\pv(t_{i+1};\pv,\xv,t_i)$ and $\xv\leftarrow\xv(t_{i+1};\pv,\xv,t_i)$, and then $i\leftarrow i+1$.
    \item In the event of scattering: Values for $(\iv,\jv)\sim P_{\iv\jv}$ are chosen randomly, distributed according to the values of the transition probability $P_{\iv\jv}$. Then $s$ is updated to $s\leftarrow s\cdot \gamma t_{i-1}|\alpha| {\rm sign}(\alpha_{\iv\jv})$. %$s\leftarrow s\cdot \sigma\gamma t_{i-1}|\alpha| sign(\alpha_{\iv\jv})$, where $\sigma$ is equal to the number of summands in the kernel, where $\alpha_{\iv\jv}\neq 0$. 
    The factor $t_{i-1}$ comes from the length of the time integral. Finally, $\pv\leftarrow\pv+\iv\Delta P,\xv\leftarrow\xv+\jv\Delta X$. If $i<n$, jump to step 3.
    \item The trajectory is followed backward in the time interval $(0,t_n)$, i.e., $\pv\leftarrow\pv(0;\pv,\xv,t_n)$ and $\xv\leftarrow\xv(0;\pv,\xv,t_n)$, where $(\pv,\xv)$ is equal to the phase space point $(\pv_n(0),\xv_n(0))$, see Figure~\ref{fig:trajectoryBW}.
    \item $f_{{\rm w}_0}(\pv,\xv)$ is evaluated at the final position $(\pv,\xv)=(\pv_n(0),\xv_n(0))$ and $A_n\leftarrow A_n+s\e^{-T\gamma}f_{{\rm w}_0}(\pv,\xv)A(\pv_T,\xv_T,T)/P(\pv_T,\xv_T)$. If $j<N_n$, set $j\leftarrow j+1$ and jump to step 2.
    \item $n\leftarrow n+1,j\leftarrow 1$, and the algorithm jumps to step 2, unless $n=N$. In this case, the next step is executed.
    \item Finally, return $\sum_{n=0}^N A_n/N_n$.
    %Finally, $A$ is evaluated as $A\leftarrow \sum_{n=0}^N A_n/N_n$.
\end{enumerate}
\end{algorithm}

\subsection{Forward algorithm} %{Forward parameterization}
Finally, a Monte Carlo algorithm is presented, where the number of scattering events is not predetermined and the scattering times are exponentially distributed. We will use forward parameterization in this case.
%\subsubsection{Distribution of the random points}
Again, $X_n$ and $P_{X_n}$ have to satisfy the condition $\langle A\rangle_n(T)=E[X_n]$. The arguments that are randomly chosen are the same as before. The transition probability $P_{\iv\jv}$ and the density function $P$ remain the same.
% , but the density function $P$ for the initial points $\pv,\xv$ should be chosen proportionally to $|f_{{\rm w}_0}|$ if possible. 
For the scattering times $t_1,\dots,t_n$, we evaluate the joint density of the number of scattering events $n$ happening in the interval $[0,T]$, and the consecutive scattering times $(t_i)_{i\in\{1,\dots,n\}}$. Considering an exponential distribution, the density for a single scattering event is given by $\gamma\e^{-\gamma t}$. The joint density is equal to the density of the first $n$ events multiplied by the probability that the next event happens after T, which yields
\begin{eqnarray}
    \eqalign{p\big((t_i)_{i\in\{1,\dots,n\}},n\big)&=\gamma^n\prod_{i=1}^n\left(\e^{-\gamma(t_i-t_{i-1})}\right)\int_T^\infty\gamma\e^{-\gamma(t_{n+1}-t_n)}\rmd t_{n+1}\\
    &=\gamma^n\e^{-\gamma t_n}\e^{\gamma t_n}\int_T^\infty\gamma\e^{-\gamma t_{n+1}}\rmd t_{n+1}\\
    &=\gamma^n\e^{-\gamma T},}
    \label{eq:pjoint}
\end{eqnarray}
assuming $t_0=0$. This conveniently coincides with the prefactor in (\ref{Anfw}).

Combining all probability functions gives $P_{X_n}=\prod_{k=1}^n(|\alpha_{\iv_k\jv_k}|)|\alpha|^{-n}P(\pv,\xv)\gamma^n\e^{-\gamma T}$. By using the condition $\langle A\rangle_n(T)=E[X_n]$ and the result of $\langle A\rangle_n(T)$ in (\ref{Anfw}), we can evaluate the random variable as
\begin{equation}
    X_n=|\alpha|^n\frac{f_{{\rm w}_0}(\pv,\xv)}{P(\pv,\xv)}\prod_{k=1}^n\big({\rm sign}(\alpha_{\iv_k\jv_k})\big)A\big(\pv_n'(T),\xv_n'(T),T\big).
\end{equation}
% The terms $\langle A\rangle_n(T)$ are supposed to be equal to the expected value $E[X_n]$ of the random variable $X_n$. Considering all probability functions, we get
% \begin{equation}
%     \eqalign{\langle A\rangle_n(T)&=E[X_n]=\int \rmd\pv\int \rmd\xv \int_0^T dt_1\int_{t_1}^T dt_2\dots\int_{t_{n-1}}^T dt_n P(\pv,\xv)\\
%     &\cdot p\big((t_i)_{i\in\{1,\dots,n\}},n\big)\sum_{\iv_1,\jv_1}\sum_{\iv_2,\jv_2}\dots\sum_{\iv_n,\jv_n}\prod_{k=1}^n\big(P_{\iv_k\jv_k}\big)X_n.}
% \end{equation}
% By using (\ref{eq:pjoint}) and the result of $\langle A\rangle_n(T)$ in (\ref{Anfw}), we can evaluate the random variable as

\begin{figure}[h]
\begin{center}
\begin{tikzpicture}[scale=0.55, transform shape]
\node (start) [startstop2] {Start};
\node (step1) [process2, below=0.5cm of start] 
{Initialization: $M,A\leftarrow 0,j\leftarrow 1$
};
\node (step2) [process2, below=0.5cm of step1] 
{$t_i\leftarrow 0, s\leftarrow 1,$\\$\pv,\xv\sim P(\pv,\xv)$\\$\pv_0\leftarrow\pv,\xv_0\leftarrow\xv$};
\node (step3) [process2, below=0.5cm of step2] {$r\sim U(0,1),t'\leftarrow-\frac{\ln{r}}{\gamma}$};
\node (iEq0) [decision2, below=0.5cm of step3] {$t_i+t'>T$};
\node (step4) [process2, left=2.1cm of iEq0] {$\pv\leftarrow\pv'(t_i+t';\pv,\xv,t_i),$\\$\xv\leftarrow\xv'(t_i+t';\pv,\xv,t_i)$};
\node (step5) [process2, left=1cm of step3] {$\iv,\jv\sim P_{\iv\jv},$\\$s\leftarrow s\cdot |\alpha|{\rm sign}(\alpha_{\iv\jv}),$\\$\pv\leftarrow \pv-\iv\Delta P,$\\$\xv\leftarrow\xv-\jv\Delta X$\\$t_i\leftarrow t_i+t'$};
\node (step6) [process2, right=2cm of step1] {$\pv\leftarrow\pv'(T;\pv,\xv,t_i),$\\$\xv\leftarrow\xv'(T;\pv,\xv,t_i)$};
\node (step7a) [process2, below=0.5cm of step6] {$A\leftarrow A+sA(\pv,\xv,T)\frac{f_{{\rm w}_0}(\pv_0,\xv_0)}{P(\pv_0,\xv_0)}$};
\node (jEqM) [decision2, below=0.5cm of step7a] {$j=M$};
\node (step7b) [process3, right=1cm of jEqM] {$j\leftarrow j+1$};
\node (stop) [startstop2, below=0.5cm of jEqM] {Return $\leftarrow\frac{A}{M}$};

\draw [->] (start) -- (step1);
\draw [->] (step1) -- (step2);
\draw [->] (step2) -- (step3);
\draw [->] (step3) -- (iEq0);
% \draw [->] (iEq0) -- node[anchor=west] {yes} (step2);
\draw [->] (iEq0) -- node[anchor=south] {no}(step4);
\draw [->] (step4) -- (step5);
\draw [->] (step5) -- (step3);
\draw [->] (iEq0.east) -- node[anchor=south] {yes} +(2cm,0) |- (step6);
% \draw [->] (step5) -- (endPos);
% \draw [->] (endPos) -- node[anchor=south] {no} (step4);
% \draw [->] (step4.north) |- (step5.east);
% \draw [->] (endPos) -- node[anchor=west] {yes} (step6);
\draw [->] (step6) -- (step7a);
\draw [->] (step7a) -- (jEqM);
\draw [->] (jEqM) -- node[anchor=south] {no} (step7b);
\draw [->] (jEqM) -- node[anchor=west] {yes} (stop);
% \draw [->] (step7b.east) -| +(1cm,-4cm) -- +(-25cm,-4cm) |- (step2.west);
\draw [->] (step7b.south) |- +(-21.5cm,-3.5cm) |- (step2.west);
\end{tikzpicture}
\end{center}
\caption{Flow chart of the forward algorithm}
    \label{fig:flowchartFW}
\end{figure}
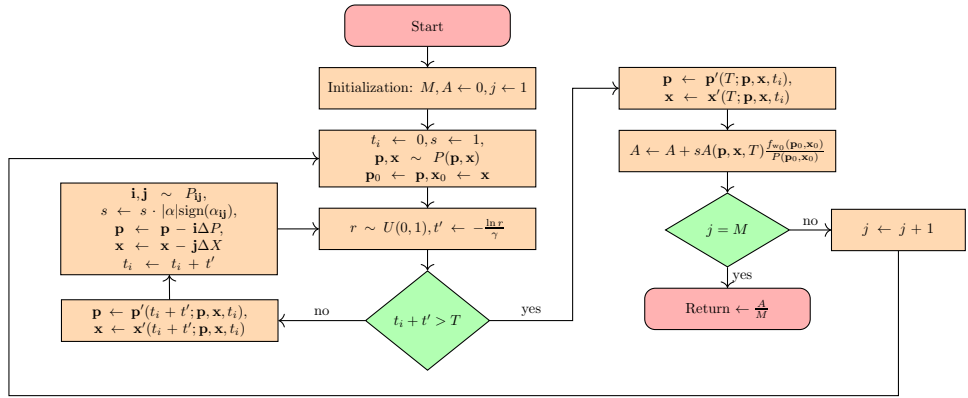

%\subsubsection{Forward trajectory algorithm for a physical quantity}
%Now that we have evaluated the random variable $X_n$ for exponentially distributed scattering times, we can develop an algorithm.
%Algorithm~\ref{algorithm:algoExpFW} describes the forward % trajectory parameterized algorithm for a physical quantity based on the derived evaluation of the random variable $X_n$ for exponentially distributed scattering times. 
The expectation value of a physical quantity can be obtained by Algorithm~\ref{algorithm:algoExpFW} (see also Figure~\ref{fig:flowchartFW}).
\begin{algorithm}{Forward algorithm}
\label{algorithm:algoExpFW}
\begin{enumerate}[label=\arabic*]
    \item Initialization of $M$ and a variable, say $A\leftarrow0$. $M$ sets the total number of the independent numerical trajectories and $A$ represents the expectation value of the physical quantity. $j\leftarrow 1$ is a counter for $M$.
    \item $t_i$ is initialized as $t_i\leftarrow 0$. $s\leftarrow1$ represents all factors in $X_n$ that are updated at each scattering event. $(\pv,\xv)\sim P(\pv,\xv)$ %, representing the initial point $\pv_0$ and $\xv_0$ for the trajectories $\pv_n'(t)$ and $\xv_n'(t)$, respectively, 
    are chosen randomly, distributed according to the chosen probability function $P(\pv,\xv)$. Since $(\pv,\xv)$ will change in the following steps, the initial values $\pv_0\leftarrow\pv,\xv_0\leftarrow\xv$ are also saved as they are needed at a later step.
    \item An exponentially distributed variable $t'\sim {\rm Exp}(\gamma)$ with the constant $\gamma$ is chosen by generating a uniformly distributed variable $r\sim {\rm U}(0,1)$ and setting $t'\leftarrow
    -\ln(r)/\gamma$. If $t_i+t'>T$, then we jump to step 6.
    \item Starting from the current $\pv$ and $\xv$ the trajectory is followed until it reaches the next scattering event at $t_i+t'$, i.e., $\pv\leftarrow\pv'(t_i+t';\pv,\xv,t_i)$ and $\xv\leftarrow\xv'(t_i+t';\pv,\xv,t_i)$.
    \item In the event of scattering: Values for $(\iv,\jv)\sim P_{\iv\jv}$ are chosen randomly, distributed according to the values of the transition probability $P_{\iv\jv}$, defined in Section~\ref{distribution}. Then, $s$ is updated to $s\leftarrow s\cdot |\alpha|{\rm sign}(\alpha_{\iv\jv})$. %$s\leftarrow s\cdot \sigma|\alpha|sign(\alpha_{\iv\jv})$, where $\sigma$ is equal to the number of summands in the kernel, where $\alpha_{\iv\jv}\neq 0$. 
    Finally, $\pv\leftarrow\pv-\iv\Delta P,\xv\leftarrow\xv-\jv\Delta X$ and $t_i\leftarrow t_i+t'$. Then jump to step 3.
    \item The trajectory is followed in the time interval $(t_i,T)$, i.e., $\pv\leftarrow\pv'(T;\pv,\xv,t_i)$ and $\xv\leftarrow\xv'(T;\pv,\xv,t_i)$, where $(\pv,\xv)$ is equal to the phase space point $(\pv_n'(T),\xv_n'(T))$, see Figure~\ref{fig:trajectoryFW}.
    \item $A(\pv,\xv,T)$ is evaluated at the final position $(\pv,\xv)=(\pv_n'(T),\xv_n'(T))$ and $A\leftarrow A+sA(\pv,\xv,T)f_{{\rm w}_0}(\pv_0,\xv_0)/P(\pv_0,\xv_0)$. $j\leftarrow j+1$.
    \item Jump to step 2, unless $j=M$. In this case, the next step is executed.
    \item Finally, return $A/M$.
\end{enumerate}
\end{algorithm}

\section{Discussion}
\label{discussionSection}
% \begin{itemize}
%     \item summary of the key findings of the work
%     \item discussion of the number of independent trajectories
%     \item possible use cases
%     \item limitation of the maximum evaluation time
%     \item extending evaluation time with generation and annihilation algorithm
%     \item little research on Wigner formalism for em fields
%     \item limitation to linear electromagnetic fields
%     \item next step magnetic field with step-wise functions
%     \item different behaviour bw $\Longleftrightarrow$
% \end{itemize}
%The analysis of the interplay between the high-order derivative term and the Liouville operator is a step in understanding gauge-invariant Wigner theory using classical Boltzmann concepts.
The two introduced Monte Carlo algorithms % are a result of an analysis of the interplay between the high-order derivative term and the Liouville operator and
constitute an important step in understanding gauge-invariant Wigner theory using classical Boltzmann concepts.
The choice of linear electromagnetic fields ensures the appearance of the same Liouville operator in both transport descriptions and thus provides a convenient reference frame for insights into the quantum evolution in terms of particles.
% It lays the foundation for simulations with a large linear factor of the magnetic field. This loosens the tight restrictions on the magnetic field. The next step could be the analysis of a magnetic field with step-wise functions. 
% IDEA: FEM triangle mesh with border condition -> use linearity.
% The solution found in this work has several similarities to an ensemble of classical electrons, governed by the Lorentz force of the electromagnetic field. The Newtonian trajectory corresponds to the free flight of the electrons, while the scattering can be interpreted as lattice vibrations. The major difference is that the Wigner function also allows non-negative values. 
The two algorithms are derived by the application of established Monte Carlo approaches for integrating the backward or forward form of the gauge-invariant Wigner equation. In the former case, the algorithm is more formal as the evolution proceeds backward in time. 
It offers computational advantages when the solution is needed locally in the phase space. Furthermore, it allows us to gradually introduce concepts used in the forward algorithm, which completes the particle picture conjectured at the end of the previous section. The quantum evolution resembles to a large extent the evolution of classical Boltzmann particles. 
An ensemble of particles is initialized in both cases according to the initial condition. Particles are accelerated by the Lorentz force over Newtonian trajectories and interrupted by scattering events. 
Comparing both algorithms reveals the proper interpretation of the distribution of the scattering times. In the backward algorithm, the scattering times were chosen uniformly distributed on the interval between the beginning of the evaluation and the previous scattering event. As a result, the scattering events tend to be unevenly distributed throughout the evolution time. The distribution density of the scattering events is inversely proportional to the length of the time intervals $(0,t_i)_{i\in\{1,\dots,n-1\}}$, and is thus higher at $0$ and lower toward $T$.
% In the beginning, they are crowded but get more and more scarce toward $T$. 
In the forward algorithm, the scattering events are evenly distributed on the interval $[0,T]$, due to the exponential distribution. This manifests in the joint probability density, which corresponds exactly to the prefactor of the terms in the resolvent expansion. 
As for the weights of the statistical sum of the physical quantity, their absolute value is multiplied by $|\alpha|$ for every scattering event. This factor corresponds to the weighted amount of possible directions the particle could scatter. Also, the sign of the weights can change during the scattering, depending on the sign of the corresponding coefficient $\alpha_{\iv\jv}$ in the kernel.

These considerations can be summarized as follows: The distribution of  scattering times is given by the formally introduced quantity $\gamma$, (\ref{gamma}), which now has been provided with a physical meaning of a total out-scattering rate  in a striking analogy with the classical counterpart.  Similarly to the latter, $\gamma$ is given by the sum of the quantities $|\alpha_{\iv\jv}|$, which corresponds to the probability for scattering from different classical mechanisms such as phonons and impurities. 
The difference  is that the terms $\alpha_{\iv\jv}$ carry a sign, so that each  scattering event can change both the absolute value of the weight and the sign, which are the main attributes of a quantum particle. Indeed, in this way scattering determines the difference between classical and quantum evolution, as discussed before. 
Furthermore, while in the former case, scattering is local in space, causing only a shift in momentum, quantum scattering leads to spatial shifts. These shifts depend on the finite difference scheme, however, this is irrelevant to the conceptual understanding: Similarly, considering computational approaches, different numerical schemes can be applied to find the numerical solution.

% Comparing both algorithms reveals the proper distribution of the scattering times. Counterintuitively, if the scattering times are chosen consecutively with a uniform distribution, they tend to be crowded at the end of the evaluation and sparse at the beginning. On the other hand, for an exponential distribution, the scattering times are evenly distributed on the interval $[0,T]$. This manifests in the joint probability density, which corresponds exactly to the prefactor of the terms in the resolvent expansion.
% As for the weights of the statistical sum of the physical quantity, their absolute value is multiplied by a factor $|\alpha|$, while their sign can switch at every scattering event, based on the scattering direction.

The introduction of the Newtonian trajectory enables us to transform the gauge-invariant Wigner equation to a Fredholm integral equation, where a resolvent expansion gives an iterative solution. However, this involves the approximation of the high-order derivative term leading to many terms in the kernel. This consequently increases the number of possible paths of the trajectory %by a factor of 15 for each scattering event, i.e., this number can explode very fast.
giving rise to the accumulation of the weight of a trajectory with the evolution.
Large positive and negative  weight values need to cancel each other in the statistical estimators for the physical averages.
Thus, the maximum simulation time $T$ of the simulation is limited, because the larger $T$, the higher the impact of the terms with a higher number of scattering events.
From a computational point of view, this leads to the well-known 'sign problem' of quantum mechanics. A good example is the Taylor series of $e^{-x}$ for large positive $x$, where large terms compensate each other to give a value smaller than unity. The problem can be addressed by using the Markovian character of the evolution of the particle ensemble, which, in particular, provides the Wigner solution $f_w$ in the entire phase space: $T$ can be decomposed on shorter time intervals $\Delta t$, so that the solution at the end of the $n$-th interval  $f_w(n\Delta T)$ becomes the initial condition for the $n+1$-th interval.
% To counter this problem, an algorithm could be developed, where all possible paths of a trajectory are calculated in one iteration. This involves a generation and annihilation process, where some paths are eliminated before their number gets too large. 
% The choice of the parameterization for both algorithms was demonstrative and could be exchanged with a few adjustments. Different behavior is expected depending on the parameterization, thus it should be considered when running simulations. Both algorithms also confirm that scattering becomes obsolete if the evolution time is small compared to the mean scattering time. In the first algorithm, the zeroth term dominates if the product of the scattering times is too small. In the second algorithm, the first scattering event will exceed the evolution time almost every time.

% \section{Text for inclusion}

%However there are major differences between classical and quantum evolution. The scattering by phonons is instantaneous in time and local in space. In our model scattering is also instantaneous, but is nonlocal in space as it involves spatial transitions. Furthermore $f_{{\rm w}_0}$ can be any legitimate initial Wigner function and thus allows for negative values, in contrast to the initial Boltzmann distribution which is a probability: The weight of the quantum particles can be negative, which is easily seen already if considering the expression for $\lange A\rangle_0$: Indeed the sign of $f_{{\rm w}_0}$ takes part in the corresponding integral. In the next section we will see that

\ack
This research was funded by the Austrian Science Fund (FWF): P33609-N and P37080-N.

\section*{References}
\bibliographystyle{iopart-num}
\bibliography{paper}

\appendix

\section{\label{writtenOut}Fredholm equation}
% The written out form of the equation~(\ref{eq:fredholm}) is given by
% \begin{eqnarray}
% \eqalign{
% \fl f_{\rm w}\bigl(\pv,\xv,t\bigr)=\e^{-t\gamma}f_{{\rm w}_0}\bigl(\pv(0),\xv(0)\bigr)+\int_0^{t} dt'\gamma\e^{-(t-t')\gamma}\\
% \fl \cdot\big[4f_{\rm w}(p_x(t'), p_y(t') + \Delta P, x(t') + \Delta X, y(t'))- 8 f_{\rm w}(p_x(t'), p_y(t'), x(t') + \Delta X, y(t'))\\
% \fl  + 4f_{\rm w}(p_x(t'), p_y(t') - \Delta P, x(t') + \Delta X, y(t')) - 4f_{\rm w}(p_x(t'), p_y(t') + \Delta P, x(t') - \Delta X, y(t')) \\
% \fl + 8 f_{\rm w}(p_x(t'), p_y(t'), x(t') - \Delta X, y(t')) - 4f_{\rm w}(p_x(t'), p_y(t') - \Delta P, x(t') - \Delta X, y(t')) \\
% \fl -f_{\rm w}(p_x(t') + \Delta P, p_y(t') + \Delta P, x(t'), y(t') + \Delta X) \\
% \fl+ f_{\rm w}(p_x(t') + \Delta P, p_y(t') - \Delta P, x(t'), y(t') + \Delta X) \\
% \fl +f_{\rm w}(p_x(t') - \Delta P, p_y(t') + \Delta P, x(t'), y(t') + \Delta X) \\
% \fl- f_{\rm w}(p_x(t') - \Delta P, p_y(t') - \Delta P, x(t'), y(t') + \Delta X)  \\
% \fl +  f_{\rm w}(p_x(t') + \Delta P, p_y(t') + \Delta P, x(t'), y(t') - \Delta X) \\
% \fl- f_{\rm w}(p_x(t') + \Delta P, p_y(t') - \Delta P, x(t'), y(t') - \Delta X)   \\
% \fl  - f_{\rm w}(p_x(t') - \Delta P, p_y(t') + \Delta P, x(t'), y(t') - \Delta P) \\
% \fl+ f_{\rm w}(p_x(t') - \Delta P, p_y(t') - \Delta P, x(t'), y(t') - \Delta X)+f_{\rm w}\bigl(\pv(t'),\xv(t'),t'\bigr)\big].}
% \label{eqFDM}
% \end{eqnarray}
The detailed form of the kernel~(\ref{eq:fh1}) is given by
\begin{eqnarray}
\eqalign{
\fl\K(\pv,\xv,t,\pv',\xv',t')&=\theta(t-t')\gamma\e^{-(t-t')\gamma}\\
&\fl\cdot\big[ 4\delta(p_x(t')-p_x', p_y(t') + \Delta P-p_y', x(t') + \Delta X-x', y(t')-y') \\
&\fl- 8 \delta(p_x(t')-p_x', p_y(t')-p_y', x(t') + \Delta X-x', y(t')-y') \\
&\fl+ 4\delta(p_x(t')-p_x', p_y(t') - \Delta P-p_y', x(t') + \Delta X-x', y(t')-y')  \\
&\fl- 4\delta(p_x(t')-p_x', p_y(t') + \Delta P-p_y', x(t') - \Delta X-x', y(t')-y') \\
&\fl+ 8 \delta(p_x(t')-p_x', p_y(t')-p_y', x(t') - \Delta X-x', y(t')-y')\\
&\fl- 4\delta(p_x(t')-p_x', p_y(t') - \Delta P-p_y', x(t') - \Delta X-x', y(t')-y') \\
&\fl-\delta(p_x(t') + \Delta P-p_x', p_y(t') + \Delta P-p_y', x(t')-x', y(t') + \Delta X-y') \\
&\fl+ \delta(p_x(t') + \Delta P-p_x', p_y(t') - \Delta P-p_y', x(t')-x', y(t') + \Delta X-y')   \\
&\fl+\delta(p_x(t') - \Delta P-p_x', p_y(t') + \Delta P-p_y', x(t')-x', y(t') + \Delta X-y') \\
&\fl- \delta(p_x(t') - \Delta P-p_x', p_y(t') - \Delta P-p_y', x(t')-x', y(t') + \Delta X-y')  \\
&\fl+  \delta(p_x(t') + \Delta P-p_x', p_y(t') + \Delta P-p_y', x(t')-x', y(t') - \Delta X-y') \\
&\fl- \delta(p_x(t') + \Delta P-p_x', p_y(t') - \Delta P-p_y', x(t')-x', y(t') - \Delta X-y') \\  
&\fl- \delta(p_x(t') - \Delta P-p_x', p_y(t') + \Delta P-p_y', x(t')-x', y(t') - \Delta X-y') \\
&\fl+ \delta(p_x(t') - \Delta P-p_x', p_y(t') - \Delta P-p_y', x(t')-x', y(t') - \Delta X-y')\\
&\fl+\delta\bigl(\pv(t')-\pv',\xv(t')-\xv'\bigr) \big].}
\label{eq:kernel}
\end{eqnarray}
\section{\label{RE}Resolvent expansion}
\subsection {Fredholm integral equation of the second kind}
The solution for a general Fredholm equation of the second kind, $f(s)=f_{\rm i}(s)+\int_a^b\K(s,s')f(s')ds'$ is given by
\begin{eqnarray}
    \eqalign{\fl f(s)=&\sum_{n=0}^\infty f_n(s),\\
    \fl f_0(s):=&f_{\rm i}(s), \\
    \fl f_n(s):=&\int_a^b\dots\int_a^b\int_a^b\K(s,t_1)\K(t_1,t_2)\cdots\K(t_{n-1},t_n)f_{\rm i}(t_n)\rmd t_1\rmd t_2\dots \rmd t_n, }\label{eq:liouvilleNeumann1} 
\end{eqnarray}
provided that the series converges.
To derive the solution with forward parameterization, the adjoint integral equation of the Fredholm equation is used. In general, given a Fredholm equation problem with $\K:[a,b]\times[a,b]\rightarrow\R, f_{\rm i}:[a,b]\rightarrow\R, a,b\in[-\infty,\infty], a<b$, like in (\ref{eq:liouvilleNeumann1}), let $g_{\rm i}:[a,b]\rightarrow\R$. Then, the adjoint equation is defined as
\begin{equation}
    g(s')=g_{\rm i}(s')+\int_a^b\K(s,s')g(s)\rmd s,
\end{equation}
with a not specified initial function $g_{\rm i}$. The solution for $g$ is also given by a resolvent expansion, but the order of the variables in $\K$ is reversed, i.e.,
\begin{eqnarray}
    \eqalign{
    \fl g(s)&=\sum_{n=0}^\infty g_n(s),\\
    \fl g_0(s)&:=g_{\rm i}(s),\\
    \fl g_n(s)&:=\int_a^b\dots\int_a^b\int_a^b\K(t_1,s)\K(t_2,t_1)\cdots\K(t_{n},t_{n-1})g_{\rm i}(t_n)\rmd t_1\rmd t_2\dots \rmd t_n. }\label{eq:liouvilleNeumann4} 
\end{eqnarray}
\subsection{Exchange Lemma}
\label{adjointLemma}
% The proof for this statement is easy Bad statement
Let $f:[a,b]\rightarrow\R$ be the solution of a Fredholm equation, where $\K:[a,b]\times[a,b]\rightarrow\R, f_{\rm i}:[a,b]\rightarrow\R$, and let $g:[a,b]\rightarrow\R$ be the solution of the adjoint equation, where $g_{\rm i}:[a,b]\rightarrow\R$. Then, there holds
\begin{equation}
    \label{eq:adjointLemma}
    \int_a^bf_{\rm i}(s)g(s)\rmd s=\int_a^bf(s)g_i(s)\rmd s.
\end{equation}
\begin{proof}
    \begin{eqnarray*}
        \int_a^bf_{\rm i}(s)g(s)\rmd s&=&\int_a^bg(s)\bigg[f(s)-\int_a^b\K(s,s')f(s')ds'\bigg]\rmd s\\
        &=&\int_a^bg(s)f(s)\rmd s-\int_a^b\int_a^b\K(s,s')f(s')g(s)\rmd s'\rmd s\\
        &=&\int_a^bf(s')\bigg[g(s')-\int_a^b\K(s,s')g(s)\rmd s\bigg]\rmd s'\\
        &=&\int_a^bf(s')g_{\rm i}(s')\rmd s'.
    \end{eqnarray*}
\end{proof}

This fact can be used to express the solution $f$ by the adjoint solution $g$ at a given point $s\in[a,b]$. In particular, if we set $g_{\rm i}(s'):=\delta(s-s')$, we can show that
\begin{equation}
    \fl f(s)=\int_a^bf(s')\delta(s-s')\rmd s'=\int_a^bf(s')g_{\rm i}(s')\rmd s'=\int_a^bf_{\rm i}(s')g(s')\rmd s'.
    \label{solutionAdjoint}
\end{equation}

% \subsection{Monte Carlo Evaluation of the Iterative Integral Terms}
% Now we are going to evaluate the integrals of the terms of the iterative solutions (\ref{An}) and (\ref{Anfw}) by Monte Carlo methods. Generally, Monte Carlo methods can be used to evaluate integrals of the form
% \begin{equation}
%     E[X]=\int_\Omega X(\xv)p(\xv)d\xv,
%     \label{eq:EX}
% \end{equation}
% with a random variable $X$ and a probability density $p$. By setting $X(x):=\frac{f(x)}{p(x)}$, one can evaluate the integral $\int_a^bf(x)dx$ of an arbitrary integrable function $f:[a,b]\rightarrow\R$. The mean $\eta$ of a sample $x_1,\dots,x_n$ is then given as
% \begin{equation}
%     \eta = \frac{1}{N}\sum_{i=1}^NX(x_i)=\frac{1}{N}\sum_{i=1}^N\frac{f(x_i)}{p(x_i)}.
% \end{equation}
% The standard deviation of $\eta$ is minimal if $p$ is proportional to $|f|$. Therefore, we have to choose our probability functions cleverly to obtain a fast converging series.

% \begin{comment}
\section{ Proof for the solution of the Wigner function}
\label{prooff_w}
The solution of (\ref{eq:fh1}) is given by the series $\sum_{n=0}^\infty f_n(\pv,\xv,t)$, where
\begin{eqnarray}
\eqalign{
f_0(\pv,\xv,t)=&~\e^{-t\gamma}f_{{\rm w}_0}\big(\pv_0(0),\xv_0(0)\big)\\
f_n(\pv,\xv,t)=&~\gamma^n\e^{-\gamma t}\int_0^tdt_1\int_0^{t_1}dt_2\dots\int_0^{t_{n-1}}dt_n\\ &\sum_{\iv_1,\jv_1}\sum_{\iv_2,\jv_2}\dots\sum_{\iv_n,\jv_n}\prod_{k=1}^n(\alpha_{\iv_k,\jv_k})f_{{\rm w}_0}\big(\pv_n(0),\xv_n(0)\big).}
\end{eqnarray}
\begin{proof}
    First, we show
    \begin{eqnarray*}
        &&\fl\int d\pv_{1}\int d\xv_{1}\dots\int d\pv_{n}\int d\xv_{n}
        \delta\big(\pv(t_1)+\iv_1\Delta P-\pv_1,\xv(t_1)+\jv_1\Delta X-\xv_1\big)\\
        &&\fl\cdot\prod_{k=1}^{n-1}\delta\big(\pv(t_{k+1};\pv_k,\xv_k,t_k)+\iv_{k+1}\Delta P-\pv_{k+1},\xv(t_{k+1};\pv_k,\xv_k,t_k)+\jv_{k+1}\Delta X-\xv_{k+1}\big)\\
        &&\fl\cdot f_{{\rm w}_0}\big(\pv(0),\xv(0)\big)=f_{{\rm w}_0}\big(\pv_{n}(0),\xv_{n}(0)\big),
    \end{eqnarray*}
    by induction, assuming $(\pv,\xv,t)$ as initial point and $t>t_1>t_2>\dots>t_{n+1}$. \\
    \textbf{Base Case:}
    \begin{equation*}
        f_{{\rm w}_0}\big(\pv(0),\xv(0)\big)=f_{{\rm w}_0}\big(\pv_0(0),\xv_0(0)\big).
    \end{equation*}
    \textbf{Induction Step:}
    \begin{eqnarray*}
        &&\fl\int d\pv_{1}\int d\xv_{1}\dots\int d\pv_{n+1}\int d\xv_{n+1}
        \delta\big(\pv(t_1)+\iv_1\Delta P-\pv_1,\xv(t_1)+\jv_1\Delta X-\xv_1\big)\\
        &&\fl\cdot\prod_{k=1}^{n}\delta\big(\pv(t_{k+1};\pv_k,\xv_k,t_k)+\iv_{k+1}\Delta P-\pv_{k+1},\xv(t_{k+1};\pv_k,\xv_k,t_k)+\jv_{k+1}\Delta X-\xv_{k+1}\big)f_{{\rm w}_0}\big(\pv(0),\xv(0)\big)\\
        &&\fl=\int d\pv_{n+1}\int d\xv_{n+1}\delta\big(\pv(t_{n+1})+\iv\Delta P-\pv_{n+1},\xv(t_{n+1})+\jv\Delta X-\pv_{n+1}\big)f_{{\rm w}_0}\big(\pv_n(0),\xv_n(0)\big)\\
        &&\fl=f_{{\rm w}_0}\Big(\pv\big(0;\pv_n(t_{n+1})+\iv\Delta P,\xv_n(t_{n+1})+\jv\Delta X,t_{n+1}\big),
        % \\&&\fl~\quad~\quad~
        \xv\big(0;\pv_n(t_{n+1})+\iv\Delta P,\xv_n(t_{n+1})+\jv\Delta X,t_{n+1}\big)\Big)\\
        &&\fl=f_{{\rm w}_0}\big(\pv_{n+1}(0),\xv_{n+1}(0)\big),
    \end{eqnarray*}
    where we have used (\ref{eq:XnPnBw}). Using this result in the resolvent expansion (\ref{eq:liouvilleNeumann1}) yields the desired result.
\end{proof}

% \end{comment}
%\section{\label{proof} Proof for the Solution of the Adjoint Equation}

% \nocite{*}

\end{document}